\advance\hoffset by 5mm
\documentclass[epj]{svjour}
\let\makeheadbox\relax

\usepackage{amsmath,amssymb}
\usepackage{feynmp}
\newcommand{\braket}[1]{\left\langle #1 \right\rangle }
\newcommand{\Tprod}[1]{\mathop{\rm T}\left[#1\right]}
\newcommand{\Greensfunc}[1]{\braket{0\left|\Tprod{#1}\right|0}}
\DeclareMathOperator{\tr}{tr}
\renewcommand{\Im}{\text{Im}}
\newcommand{\dd}{\mathrm{d}}
\newcommand{\ii}{\mathrm{i}}
\allowdisplaybreaks
\numberwithin{equation}{section}
\begin{document}
\begin{fmffile}{nlhpics}
\fmfset{arrow_len}{2mm}
\title{Consistent Dyson summation of Higgs propagators in nonlinear parameterizations revisited}
\author{ Christian Schwinn\thanks{schwinn@thep.physik.uni-mainz.de}}
\institute{Institut f\"ur Physik, Johannes-Gutenberg-Universit\"at, Staudingerweg 7,  D-55099 Mainz, Germany  } 
\date{July 5, 2004, revised November 17, 2004}
\headnote{\normalsize{MZ-TH/04-06\\  hep-ph/0407053}}
 \abstract{
  As we demonstrate in a process independent way, in a nonlinear
  parameterization of the scalar sector of the standard model the Dyson
  summation of the Higgs self energy can be performed without violating the
  Ward Identities. This implies also the Goldstone boson
  equivalence theorem, in the limited range of its validity in effective
  field theories.  This
  proves an earlier conjecture of Valencia and Willenbrock.  Furthermore, the
  full Higgs propagator is independent of the gauge parameters.  These results
  are consistent with the extension of the `gauge flip' formalism for the
  construction of gauge invariant classes of Feynman diagrams to loop
  diagrams.  In a nonlinear parameterization of a 2-Higgs doublet model, the
  consistent Dyson summation is possible for all neutral Higgs bosons, but not
  for the charged scalars.  Explicit examples of the equivalence theorem are
  discussed both in the minimal standard model and a two-Higgs doublet
  model.
}
\maketitle
\section{Introduction}
Despite the phenomenological success of the electroweak standard
model~(SM), the underlying mechanism of electroweak symmetry breaking
remains to be verified experimentally. Assuming this symmetry breaking
is implemented in nature by the Higgs mechanism, one of the main goals
of the next generation of collider experiments will be the study of
the Higgs sector~\cite{Carena:2002es}.  An extended Higgs sector is
predicted for instance by the minimal supersymmetric standard model
that involves a specific two Higgs doublet model~(2HDM). 
The determination of the couplings of the Higgs bosons makes it necessary
to consider processes with up to six or eight fermions in the final state
for which a complete calculation of radiative corrections is currently
not viable.

An useful
tool to probe the symmetry breaking sector of the SM is the Goldstone
boson equivalence theorem
(ET)~\cite{Chanowitz:1985,Yao:1988,Veltman:1990,Dobado:1994} that
relates scattering amplitudes of longitudinally polarized massive
gauge bosons to that of the associated Goldstone bosons~(GBs).  In the
heavy Higgs limit, one can further replace internal gauge boson
lines by GBs~\cite{Marciano:1988un}, allowing to simplify higher order
calculations of heavy Higgs effects~\cite{Maher:1993vj}.
 
Calculations of cross sections involving resonant Higgs bosons require a
careful treatment of the Higgs-resonance, especially since the width depends
strongly on the Higgs mass.  As is well known for massive gauge bosons, the
violations of gauge invariance from inconsistent prescriptions for finite
width effects can lead to dramatic errors in the calculations of cross
sections.  It has been shown that fermionic self energy contributions can be
resummed consistently~\cite{Argyres:1995} if the fermionic corrections to
irreducible vertices are evaluated at the same order of perturbation theory.
For the Higgs self energy, however, the fermionic contributions will be not
sufficient in the mass region $m_H> 2m_W$ where decay into gauge bosons gives
the dominant contribution to the the Higgs width.  A consistent treatment of
the bosonic contributions is possible in the framework of the background field
gauge~\cite{Abbott:1981hw} that requires a calculation of the complete
radiative corrections at a fixed loop order which is presently not viable for
the multi-fermion final states relevant for the measurement of the Higgs boson
couplings. While a dependence on the quantum gauge parameter remains, in the
Feynman gauge the background field gauge reproduces the results of the pinch
technique~\cite{Cornwall:1989gv}.  Other suggested schemes for the treatment
of unstable particles include the pole scheme~\cite{Stuart:1991xk} and the use
of an effective Lagrangian including Wilson lines~\cite{Beenakker:1999}.
Recently an approach based on collinear effective field theory has been
proposed~\cite{Beneke:2003xh} that has not yet been applied to realistic
calculations.  For suggested schemes for the treatment of the Higgs resonance
see
e.g.~\cite{Valencia:1990jp,Seymour:1995qg,Denner:1996ug,Papavassiliou:1998pb,Kilian:1998bh}.

The compatibility of the Dyson summation of the Higgs propagator with
the equivalence theorem has
been discussed by Valencia and Willenbrock~\cite{Valencia:1990jp}.
Since the ET holds order by order in perturbation theory while the
Dyson summation mixes different orders, one cannot expect that the ET
holds in a `naive' way when finite widths are introduced.  Indeed, as
has been demonstrated in~\cite{Valencia:1990jp}, a careful treatment
of vertex corrections and non resonant contributions is necessary to
establish the ET in the Higgs resonance region.  It was conjectured
in~\cite{Valencia:1990jp} that the situation is simpler in a nonlinear
parameterization of the SM~\cite{Appelquist:1980vg} and a `naive'
version of the ET is satisfied i.e. the gauge boson and the Goldstone
boson amplitude agree manifestly, also after introduction of a finite
Higgs width and without taking vertex corrections into account.  It is
plausible that such a simplification occurs in a nonlinear
parameterization where the Higgs boson transforms trivially under
gauge transformations and Becchi-Rouet-Stora-Tuytin~(BRST)
transformations, allowing to construct effective field theories
without~\cite{Appelquist:1980vg} or with a nonstandard Higgs
boson~\cite{Chivukula:1993,Han:2000ic}.  Thus one can expect that Feynman
diagrams involving Higgs bosons can be consistently treated separately
from those without Higgs bosons.  Indeed, using the `gauge flip'
formalism for the construction of gauge invariant classes~(GICs) of
tree level Feynman diagrams~\cite{Boos:1999}, a classification of GICs
of tree level diagrams in the nonlinearly parameterized SM~(NL-SM) in
terms of the number of internal Higgs bosons has been given
in~\cite{OS:FGH}.

In a nonlinear parameterization, such a decomposition of the amplitude
need not respect the good high energy behaviour implied by partial
wave unitarity, despite being gauge invariant.  Also the applicability
of the ET is limited~\cite{Dobado:1994} compared to linear
parameterizations.  Nevertheless, disentangling the Higgs diagrams
from the more involved gauge boson contributions and using the ET to
work with the simpler GB amplitudes allows a more transparent
discussion of the unitarity violations induced by a finite Higgs
width. This has been used in~\cite{Seymour:1995qg} to obtain a simple
prescription for the Higgs propagator including a running width
without violating unitarity in gauge boson scattering. Such a
prescription is applicable in tree level calculations and can be
implemented in computer programs for the generation of scattering
matrix elements.  The prescription of~\cite{Seymour:1995qg} has also
proven useful to obtain unitarity bounds on couplings of a
non-standard Higgs boson~\cite{Han:2000ic}.

In this note, we revisit the properties of Higgs propagators in
nonlinearly parameterized scalar sectors, providing a first
example for the conjectured extension of the `flip' formalism 
to loop diagrams~\cite{Boos:1999}.
While a formal proof of this formalism on the one loop level will
be given elsewhere~\cite{Ondreka:2003,OOS:prep}, 
in this note we use it as an intuitive tool and give
direct proofs for the properties suggested by this formalism. 

In section~\ref{sec:groves} we review the formalism of
flips and its extension to loop diagrams~\cite{Ondreka:2003,OOS:prep}.
Applied to the Higgs resonance in the NL-SM, the flip formalism is
consistent with the conjecture of~\cite{Valencia:1990jp} and, in
addition, with gauge parameter independence of the Higgs propagator.
In section~\ref{sec:et} we discuss the ET in the NL-SM and the
high energy behaviour of indvividual GICs.  In section~\ref{sec:nl}
we give a formal proof and a one loop example for gauge parameter
independence of the Higgs propagator in the NL-SM, including a
discussion of $H$-$Z$ mixing induced by CP violation.  In
section~\ref{sec:2hdm} we extend our analysis to a nonlinear parameterization
of the 2HDM~\cite{Ciafaloni:1997ur}.  We show that a
consistent Dyson summation can only be performed for the neutral Higgs
bosons, including the CP odd scalar while it is inconsistent for the
charged Higgs bosons.  We give an explicit example for the violation
of the naive ET by the width of the charged Higgs bosons.

\section{Flips, Groves and the Higgs resonance}\label{sec:groves}
\subsection{Gauge invariant classes of tree  diagrams}
Let us briefly review the formalism of~\cite{Boos:1999} for the determination
of GICs of tree diagrams and the application to nonlinearly parameterized
scalar sectors~\cite{OS:FGH}, before we describe its extension to loop
diagrams and implications for the description of the Higgs resonance in
subsection~\ref{sec:loop-flips}.  GICs are defined as subsets of Feynman
diagrams contributing to a scattering matrix element that satisfy the
appropriate Ward Identities~(WIs) by themselves and are gauge parameter
independent. GICs can also be defined for Green's functions with off-shell
particles~\cite{OS:FGH} but the relevant definition involving the Slavnov
Taylor Identities~(STIs) is rather technical and will not be reproduced here.

In ~\cite{Boos:1999} it
 has been realized that the problem
of constructing  GICs of tree diagrams can be solved recursively by
considering minimal sets of 4-point sub-diagrams that satisfy the
appropriate WIs. In an unbroken gauge theory they are given by:
 \begin{subequations}
\label{subeq:gauge_flips}
\begin{align}
\label{eq:gauge_flips2}
G_{4,2F} &= 
\left\{
\parbox{15\unitlength}{
\begin{fmfgraph}(15,15)
\fmfleft{a,b}
\fmfright{f1,f2}
\fmf{fermion}{a,fwf1}
\fmf{fermion}{fwf1,fwf2}
\fmf{fermion}{fwf2,b}
\fmf{photon}{fwf1,f1}
\fmf{photon}{fwf2,f2}
\fmfdot{fwf1,fwf2}
\end{fmfgraph}}\,,\,
\parbox{15\unitlength}{
\begin{fmfgraph}(15,15)
\fmfleft{a,b}
\fmfright{f1,f2}
\fmf{fermion}{a,fwf1}
\fmf{fermion}{fwf1,fwf2}
\fmf{fermion}{fwf2,b}
\fmf{phantom}{fwf2,f2}
\fmf{phantom}{fwf1,f1}
\fmffreeze
\fmf{photon}{fwf2,f1}
\fmf{photon,rubout}{fwf1,f2}
\fmfdot{fwf1,fwf2}
\end{fmfgraph}}\,,\, 
\parbox{15\unitlength}{
\begin{fmfgraph}(15,15)
\fmfleft{a,b}
\fmfright{f1,f2}
\fmf{fermion}{a,fwf}
\fmf{fermion}{fwf,b}
\fmf{photon}{fwf,www}
\fmf{photon}{www,f1}
\fmf{photon}{www,f2}
\fmfdot{www,fwf}
\end{fmfgraph}}\right \}\\
G_4 &=
\left\{
\parbox{15\unitlength}{
\begin{fmfgraph}(15,15)
\fmfleft{a,b}
\fmfright{f1,f2}
\fmf{photon}{a,fwf1}
\fmf{photon}{fwf1,fwf2}
\fmf{photon}{fwf2,b}
\fmf{photon}{fwf1,f1}
\fmf{photon}{fwf2,f2}
\fmfdot{fwf1}
\fmfdot{fwf2}
\end{fmfgraph}}\,,\,
\parbox{15\unitlength}{
\begin{fmfgraph}(15,15)
\fmfleft{a,b}
\fmfright{f1,f2}
\fmf{photon}{a,fwf1}
\fmf{photon}{fwf1,fwf2}
\fmf{photon}{fwf2,b}
\fmf{phantom}{fwf2,f2}
\fmf{phantom}{fwf1,f1}
\fmffreeze
\fmf{photon}{fwf2,f1}
\fmf{photon,rubout}{fwf1,f2}
\fmfdot{fwf1}
\fmfdot{fwf2}
\end{fmfgraph}}\, ,\, 
\parbox{15\unitlength}{
\begin{fmfgraph}(15,15)
\fmfleft{a,b}
\fmfright{f1,f2}
\fmf{photon}{a,fwf}
\fmf{photon}{fwf,b}
\fmf{photon}{fwf,www}
\fmf{photon}{www,f1}
\fmf{photon}{www,f2}
\fmfdot{fwf}
\fmfdot{www}
\end{fmfgraph}}\, , \, 
\parbox{15\unitlength}{
\begin{fmfgraph}(15,15)
\fmfleft{a,b}
\fmfright{f1,f2}
\fmf{photon}{a,c}
\fmf{photon}{c,b}
\fmf{photon}{c,f1}
\fmf{photon}{c,f2}
\fmfdot{c}
\end{fmfgraph}}\right \}
\end{align}
\end{subequations}
The construction of GICs is based on defining elementary
`gauge flips' as exchanges of diagrams in these sets. 
Gauge flips among larger diagrams are defined 
by applying elementary flips to sub-diagrams.
For instance, 
the  five point diagrams
\begin{multline}\label{eq:5point-grove}
\parbox{20mm}{
 \begin{fmfgraph}(20,15)
\fmftopn{t}{2}
\fmfbottomn{b}{2}
\fmfright{r}
\fmf{fermion}{t1,t}
\fmf{fermion}{t,b1}
\fmf{photon}{v2,t}
\fmf{fermion,tension=2}{v1,t2}
\fmf{plain,tension=2}{v1,v2}
\fmf{fermion}{b2,v2}
\fmffreeze
\fmf{photon}{r,v1}
\fmfdot{v1,v2,t}
\end{fmfgraph}}\, \Leftrightarrow\, 
\parbox{20\unitlength}{
\begin{fmfgraph}(20,15)
\fmftopn{t}{2}
\fmfbottomn{b}{2}
\fmfright{r}
\fmf{fermion}{t1,t,b1}
\fmf{photon}{v1,t}
\fmf{fermion}{v1,t2}
\fmf{plain,tension=2}{v1,v2}
\fmf{fermion,tension=2}{b2,v2}
\fmffreeze
\fmf{photon}{r,v2}
\fmfdot{v1,v2,t}
\end{fmfgraph}} \Leftrightarrow
\parbox{20\unitlength}{
\begin{fmfgraph}(20,15)
\fmftopn{t}{3}
\fmfbottomn{b}{2}
\fmfright{r}
\fmf{fermion}{b2,t,t3}
\fmf{photon,tension=2}{v1,v2}
\fmf{photon,tension=2}{v2,t}
\fmf{fermion}{t1,v1,b1}
\fmffreeze
\fmf{photon}{t2,v2}
\fmfdot{v1,v2,t}
\end{fmfgraph}}\\
\Leftrightarrow \,
\parbox{20\unitlength}{
\begin{fmfgraph}(20,15)
\fmfbottomn{b}{2}
\fmftopn{t}{3}
\fmf{fermion}{b2,t,t3}
\fmf{photon}{v2,t}
\fmf{fermion,tension=2}{t1,v1}
\fmf{plain,tension=2}{v1,v2}
\fmf{fermion}{v2,b1}
\fmffreeze
\fmf{photon}{t2,v1}
\fmfdot{v1,v2,t}
\end{fmfgraph}}\,\Leftrightarrow \,
\parbox{20\unitlength}{
\begin{fmfgraph}(20,15)
\fmftopn{t}{2}
\fmfbottomn{b}{3}
\fmfright{r}
\fmf{fermion}{b3,t,t2}
\fmf{photon}{v1,t}
\fmf{fermion}{t1,v1}
\fmf{plain,tension=2}{v1,v2}
\fmf{fermion,tension=2}{v2,b1}
\fmffreeze
\fmf{photon}{b2,v2}
\fmfdot{v1,v2,t}
\end{fmfgraph}}
\end{multline}
are connected by elementary gauge flips, as denoted by a double arrow.
This procedure of `flipping' gauge bosons through diagrams is in fact
a formalization of the prescription to insert a gauge boson at all
possible places into a diagram, familiar from the diagrammatic proof
of the WI in QED. It should be noted, however, that the gauge flips
have to be applied both to external and internal gauge bosons present
in Feynman diagrams.  A set of diagrams like~\eqref{eq:5point-grove}
that is closed under the application of flips---i.e.  all diagrams in
the set are connected by gauge flips and no diagram outside the set
can be obtained by a gauge flip---is called a `grove'.  As was shown
in~\cite{Boos:1999}, groves are the \emph{minimal} GICs of tree
diagrams.  Examples for the structure of groves in the electroweak SM
can be found in~\cite{Boos:1999,OS:FGH}.

To apply the flip-formalism to the Higgs resonance, we need the the
correct form of the gauge flips in spontaneously broken gauge
theories.  If the scalar sector of the SM is parameterized linearly,
it turns out that Higgs boson exchange diagrams have to be included in
the gauge flips in addition to~\eqref{subeq:gauge_flips} while they
can be omitted in a nonlinear parameterization~\cite{OS:FGH}.
Therefore a flip 
\begin{multline}\label{eq:higgs-gauge-flip}
\left\{
\parbox{15\unitlength}{
\begin{fmfgraph}(15,15)
\fmfleft{a,b}
\fmfright{f1,f2}
\fmf{plain}{a,fwf1}
\fmf{plain}{fwf1,fwf2}
\fmf{plain}{fwf2,b}
\fmf{photon}{fwf1,f1}
\fmf{photon}{fwf2,f2}
\fmfdot{fwf1,fwf2}
\end{fmfgraph}}\,,\,
\parbox{15\unitlength}{
\begin{fmfgraph}(15,15)
\fmfleft{a,b}
\fmfright{f1,f2}
\fmf{plain}{a,fwf1}
\fmf{plain}{fwf1,fwf2}
\fmf{plain}{fwf2,b}
\fmf{phantom}{fwf2,f2}
\fmf{phantom}{fwf1,f1}
\fmffreeze
\fmf{photon}{fwf2,f1}
\fmf{photon,rubout}{fwf1,f2}
\fmfdot{fwf1,fwf2}
\end{fmfgraph}}\,,\, 
\parbox{20\unitlength}{
\begin{fmfchar*}(20,15)
\fmfleft{a,b}
\fmfright{f1,f2}
\fmf{plain}{a,fwf}
\fmf{plain}{fwf,b}
\fmf{photon,label=$Z$}{fwf,www}
\fmf{photon}{www,f1}
\fmf{photon}{www,f2}
\fmfdot{www,fwf}
\end{fmfchar*}}\right \}\\
\Leftrightarrow\,
\parbox{15\unitlength}{
\begin{fmfchar*}(20,15)
\fmfleft{l1,l2}
\fmfright{r1,r2}
\fmf{photon}{r1,fwf1}
\fmf{photon}{fwf1,r2}
\fmf{dashes,label=$H$}{fwf1,fwf2}
\fmf{plain}{fwf2,l1}
\fmf{plain}{fwf2,l2}
\fmfdot{fwf1}
\fmfdot{fwf2}
\end{fmfchar*}}
\end{multline}
is absent in the NL-SM but has to be included in a linear
parameterization.  Here and in the following, plain lines denote
arbitrary particles.  We take it as understood that in an $R_\xi$
gauge the appropriate diagrams with internal GBs have to be included
in addition.  The complete list of elementary gauge flips in nonlinear
parameterizations of general Higgs sectors can found in \cite{OS:FGH}.
We will give some details on the formal reason for this simplification
of the gauge flips at the end of this subsection.  The absence of the
flip~\eqref{eq:higgs-gauge-flip} leads to the emergence of additional
GICs in the NL-SM that can be classified according
to the number of internal Higgs bosons~\cite{OS:FGH}. This property
will also be proven using a different formalism in
section~\ref{sec:et} where it is used to proof the 
simplification of the ET observed in~\cite{Valencia:1990jp}.

As an example for the decomposition of an amplitude into groves,
consider the tree level diagrams contributing to the process $e^+
e^-\to \bar b b Z$ when the Higgs coupling to the electrons is set to
zero. In the NL-SM they fall into three groves:
\begin{subequations}
\begin{align}
G_t&=\left\{\quad
\parbox{28\unitlength}{
\fmfframe(4,4)(4,4){
\begin{fmfchar*}(20,20)
\fmftopn{t}{2}
\fmfbottomn{b}{2}
\fmfright{r}
\fmf{fermion,tension=2}{t1,v1,v2,b1}
\fmf{phantom}{v1,t2}
\fmf{photon}{v2,b2}
\fmffreeze
\fmf{photon,label=$\gamma /Z$,la.si=left}{v1,i}
\fmf{fermion}{r,i,t2}
\fmfdot{v1,v2,i}
\fmfv{label=$e^-$}{t1}
\fmfv{label=$e^+$}{b1}
\fmfv{label=$b$}{t2}
\fmfv{label=$\bar b$}{r}
\fmfv{label=$Z$}{b2}
\end{fmfchar*}}}\quad , \quad
\parbox{28\unitlength}{
\fmfframe(4,4)(4,4){
\begin{fmfchar*}(15,20)
\fmftopn{t}{2}
\fmfbottomn{b}{2}
\fmfright{r}
\fmf{fermion,tension=2}{t1,v1,v2,b1}
\fmf{photon}{v1,t2}
\fmf{phantom}{v2,b2}
\fmffreeze
\fmf{photon,label=$\gamma /Z$}{v2,i}
\fmf{fermion}{b2,i,r}
\fmfdot{v1,v2,i}
\fmfv{label=$e^-$}{t1}
\fmfv{label=$e^+$}{b1}
\fmfv{label=$b$}{r}
\fmfv{label=$\bar b$}{b2}
\fmfv{label=$Z$}{t2}
\end{fmfchar*}}}
\right\}\\
G_s&=\left\{\quad
\parbox{28mm}{
\fmfframe(4,4)(4,4){
\begin{fmfchar*}(20,20)
\fmftopn{t}{2}
\fmfbottomn{b}{2}
\fmfright{r}
\fmf{fermion}{t1,t,b1}
\fmf{photon,label=$\gamma /Z$}{v2,t}
\fmf{fermion,tension=2}{v1,t2}
\fmf{plain,tension=2}{v1,v2}
\fmf{fermion}{b2,v2}
\fmffreeze
\fmf{photon}{r,v1}
\fmfdot{v1,v2,t}
\fmfv{label=$e^-$}{t1}
\fmfv{label=$e^+$}{b1}
\fmfv{label=$b$}{t2}
\fmfv{label=$\bar b$}{b2}
\fmfv{label=$Z$}{r}
\end{fmfchar*}}}\quad , \quad 
\parbox{28\unitlength}{
\fmfframe(4,4)(4,4){
\begin{fmfchar*}(20,20)
\fmftopn{t}{2}
\fmfbottomn{b}{2}
\fmfright{r}
\fmf{fermion}{t1,t,b1}
\fmf{photon,label=$\gamma /Z$}{v1,t}
\fmf{fermion}{v1,t2}
\fmf{plain,tension=2}{v1,v2}
\fmf{fermion,tension=2}{b2,v2}
\fmffreeze
\fmf{photon}{r,v2}
\fmfdot{v1,v2,t}
\fmfv{label=$e^-$}{t1}
\fmfv{label=$e^+$}{b1}
\fmfv{label=$b$}{t2}
\fmfv{label=$\bar b$}{b2}
\fmfv{label=$Z$}{r}
\end{fmfchar*}}}\,\right\}\\
G_H&=\left\{\quad
\parbox{28\unitlength}{
\fmfframe(4,4)(4,4){
\begin{fmfchar*}(20,20)
\fmftopn{t}{2}
\fmfbottomn{b}{2}
\fmfright{r}
\fmf{fermion,tension=2}{t1,t,b1}
\fmf{photon,tension=2,label=$Z$,la.si=right}{t,v1}
\fmf{photon}{v1,b2}
\fmf{phantom}{v1,t2}
\fmffreeze
\fmf{dashes,label=$H$,la.si=left}{v1,v2}
\fmf{fermion}{r,v2,t2}
\fmfdot{v1,v2,t}
\fmfv{label=$e^-$}{t1}
\fmfv{label=$e^+$}{b1}
\fmfv{label=$b$}{t2}
\fmfv{label=$\bar b$}{r}
\fmfv{label=$Z$}{b2}
\end{fmfchar*}}}\,\right\}
\end{align}
\end{subequations}
In a linear parameterization, there is a flip connecting the
`Higgsstrahlung' diagram $G_H$ to the diagrams in $G_t$, so only two
groves remain. 

In order to be able to obtain the gauge flips in the 2HDM in
section~\ref{sec:2hdm}, we have to review the formal reason for the
simplification of the gauge flips in nonlinear parameterizations.
In a spontaneous broken gauge theory in $R_\xi$ gauge,
the diagrams that have to be taken into account in the definition of
the elementary gauge flips are determined by the requirement
that all corresponding four point diagrams with external GBs
also satisfy the appropriate WIs~\cite{OS:FGH}.  The tree level Higgs
interaction vertex functions in the NL-SM satisfy the simple WIs (here
$V_a$=$W^\pm,Z$ and $\phi_a$ are the associated GBs)
\begin{subequations}
\label{eq:h-wis}
\begin{align}
\ii p_{a\mu} \Gamma^{\mu\nu}_{V_aV_b H}+m_{V_a}\Gamma_{\phi_aV_b H}^\nu&=0
\label{eq:hww-wi}\\
\ii p_{a\mu} \Gamma^\mu_{V_a\phi_b H}+m_{V_a}\Gamma_{\phi_a\phi_bH}&=0
\label{eq:hphiw-wi}
\end{align} 
\end{subequations}
For a formal derivation of these identities and our notation used for
vertex functions see appendix~\ref{app:sti}.
Since the identities~\eqref{eq:h-wis} don't require the Higgs
to be on shell, they imply
 that the Higgs exchange diagram in~\eqref{eq:higgs-gauge-flip}
and the corresponding diagram with one external GB satisfy a WI by themselves.
Therefore the gauge flip~\eqref{eq:higgs-gauge-flip} can be omitted 
in the NL-SM. 
In a linear parameterization, there are additional terms contributing to~\eqref{eq:h-wis}
and the flip~\eqref{eq:higgs-gauge-flip} cannot be omitted. 

\subsection{Groves of loop diagrams and application to the Higgs resonance}
\label{sec:loop-flips}
The action of gauge flips can be extended to loop
diagrams~\cite{Ondreka:2003}, taking into account that the external
legs of 4-point sub-diagrams can be connected to form a closed loop.  A
computer program \texttt{mangroves} for the determination of groves of
loop diagrams is currently in preparation~\cite{Ondreka:2003}.  It is
plausible to conjecture~\cite{Boos:1999} that the groves obtained in
this way are the minimal GICs of loop diagrams, since one doesn't
expect that for loop diagrams a finer partitioning 
of the amplitude into GICs will be possible than at tree level.
Here we don't attempt a formal proof~\cite{OOS:prep} but take the
attitude that groves are sensible candidates for minimal GICs of loop
diagrams.
As an example of a grove of loop diagrams, 
consider the one loop quark gluon vertex in QCD where the application
of the flips~\eqref{subeq:gauge_flips} results in the set of diagrams
(we take it as understood that the appropriate ghost contributions have
to be included in addition)
\begin{multline}\label{eq:vertex-grove}
\parbox{20mm}{
\begin{fmfgraph*}(20,15)
\fmfleft{l1,l2}
\fmfright{r}
\fmf{fermion,tension=1}{l1,i1,v}
\fmf{fermion,tension=1}{v,i2,l2}
\fmf{photon}{v,r}
\fmffreeze
\fmf{photon,tension=0}{i1,i2}
\fmfdot{v,i1,i2}
\end{fmfgraph*}}
\quad \Leftrightarrow \\\
\left\{
 \begin{aligned} &
\parbox{20mm}{
\begin{fmfgraph*}(20,15)
\fmfleft{l1,l2}
\fmfright{r}
\fmf{fermion}{l1,v}
\fmf{fermion,tension=3}{v,i1,i2,l2}
\fmf{photon}{v,r}
\fmffreeze
\fmf{photon,left=2,tension=0}{i1,i2}
\fmfdot{v,i1,i2}
\end{fmfgraph*}}
\quad,\,
\parbox{20mm}{
\begin{fmfgraph*}(20,15)
\fmfleft{l1,l2}
\fmfright{r}
\fmf{fermion,tension=1}{l1,i1}
\fmf{photon,tension=1}{i1,v,i2}
\fmf{fermion,tension=1}{i2,l2}
\fmf{photon}{v,r}
\fmffreeze
\fmf{fermion,tension=0}{i1,i2}
\fmfdot{v,i1,i2}
\end{fmfgraph*}}\quad,
\parbox{20mm}{
\begin{fmfgraph*}(20,15)
\fmfleft{l1,l2}
\fmfright{r}
\fmf{fermion,tension=3}{l1,i1,i2,v}
\fmf{fermion}{v,l2}
\fmf{photon}{v,r}
\fmf{photon,left=2,tension=0}{i1,i2}
\fmfdot{v,i1,i2}
\end{fmfgraph*}}\\
&\parbox{20mm}{
\begin{fmfgraph*}(20,15)
\fmfleft{l1,l2}
\fmfright{r}
\fmf{fermion}{l1,v,l2}
\fmf{photon,tension=2}{v,i1}
\fmf{photon,left}{i1,i2,i1}
\fmf{photon,tension=2}{i2,r}
\fmfdot{v,i1,i2}
\end{fmfgraph*}}
\quad,\,
\parbox{20mm}{
\begin{fmfgraph*}(20,15)
\fmfleft{l1,l2}
\fmfright{r}
\fmf{fermion}{l1,v,l2}
\fmf{photon,tension=2}{v,i,r}
\fmfdot{v,i}
\fmffreeze
\fmfi{photon}{fullcircle scaled .3w shifted (vloc(__i)+(0,.15w))}
\end{fmfgraph*}}\,
\end{aligned}
\right\}
\end{multline}
The diagrams in~\eqref{eq:vertex-grove} should be considered as part
of a larger diagram so in general it will be necessary to `flip' the
gauge boson lines further through the complete diagram.  Since there
is no flip connecting the vertex correction to the fermion-loop
diagrams contributing to the vacuum polarization, these diagrams form
a separate grove.  Therefore in this example the flip-formalism is
consistent with the fermion-loop
scheme~\cite{Argyres:1995}.

Applying the flip formalism to the Higgs resonance, we immediately find 
that the Higgs self energy 
and the vertex corrections are not connected by gauge flips in the 
NL-SM because the Higgs exchange diagrams are not included in the elementary
gauge flips~\eqref{eq:higgs-gauge-flip}:
\begin{equation}
\parbox{35mm}{
\begin{fmfgraph*}(35,15)
\fmfleftn{l}{2}
\fmfrightn{r}{4}
\fmf{plain,tension=1}{l1,v,l2}
\fmf{plain,tension=2}{r4,ir1,r}
\fmf{plain,tension=2}{r1,ir2,r}
\fmf{dashes,tension=1}{v,i1}
\fmf{photon,left,tension=0.5}{i1,i2,i1}
\fmf{dashes,tension=1}{i2,r}
\fmfdot{v,r,i1,i2}
\end{fmfgraph*}}\quad\nLeftrightarrow\quad
\parbox{30mm}{
\begin{fmfgraph*}(30,15)
\fmfleftn{l}{2}
\fmfrightn{r}{4}
\fmf{plain,tension=2}{l1,i1}
\fmf{plain,tension=2}{l2,i2}
\fmf{plain,tension=2}{r1,ir1,r}
\fmf{plain,tension=2}{r,ir2,r4}
\fmf{photon,tension=1}{i1,v,i2}
\fmf{dashes,tension=1}{v,r}
\fmf{plain,tension=0}{i1,i2}
\fmfdot{v,r,i1,i2}
\end{fmfgraph*}}
\end{equation}
This property is independent of the external particles attached to the Higgs
propagators. 
Similarly, there are no flips to
irreducible higher order contributions to the self energy:
\begin{equation}\label{eq:selfenergy-flip}
\parbox{35mm}{
\begin{fmfgraph*}(35,15)
\fmfleft{l1,l2}
\fmfright{r1,r2}
\fmf{plain}{l1,v,l2}
\fmf{plain}{r1,r,r2}
\fmf{dashes,tension=0.5}{v,i1}
\fmf{photon,left,tension=0.2}{i1,i2,i1}
\fmf{dashes,tension=0.5}{i2,i3}
\fmf{photon,left,tension=0.2}{i3,i4,i3}
\fmf{dashes,tension=0.5}{i4,r}
\fmfdot{v,r,i1,i2,i3,i4}
\end{fmfgraph*}}
\quad\nLeftrightarrow\quad
\parbox{30mm}{
\begin{fmfgraph*}(30,15)
\fmfleft{l1,l2}
\fmfright{r1,r2}
\fmftop{t}
\fmfbottom{b}
\fmf{plain}{l1,v,l2}
\fmf{plain}{r1,r,r2}
\fmf{phantom,tension=3}{t,il1}
\fmf{photon,tension=1}{il1,il2}
\fmf{phantom,tension=3}{il2,b}
\fmf{dashes,tension=0.5}{v,i1}
\fmf{photon,tension=0.2}{i1,il1,i2,il2,i1}
\fmf{dashes,tension=0.5}{i2,r}
\fmfdot{v,r,i1,il1,i2,il2}
\end{fmfgraph*}}
\end{equation}
Provided the identification of groves with minimal GICs holds for loop
diagrams, the results of the flip formalism therefore indicate that a
resummation of the self energy insertions without including vertex
corrections or higher order contributions to the self energy doesn't
violate WIs or gauge parameter independence.  Since the ET is a
consequence of the WIs and the kinematical properties of the
longitudinal gauge boson polarization vector, this is consistent with
the conjecture of~\cite{Valencia:1990jp}.  In contrast, the gauge
flip~\eqref{eq:higgs-gauge-flip} has to be included in a linear parameterization so that
vertex corrections and irreducible contributions to the self energy
consistently have to be considered at the same loop order.  The gauge
parameter independence of the Higgs propagator has not been discussed
in~\cite{Valencia:1990jp} and will be treated in more detail in
section~\ref{sec:nl}.

\section{Goldstone boson equivalence theorem and the Higgs resonance}
\label{sec:et}
Using the formalism of gauge flips, we have motivated that the Dyson
summation of the Higgs resonance in the NL-SM doesn't violate WIs or
the Goldstone boson equivalence theorem, in agreement with the
conjecture of~\cite{Valencia:1990jp}.  Before providing a general
proof in subsection~\ref{sec:et-proof}, we give an explicit
example for the ET in the NL-SM and discuss the high energy behaviour of GICs
in nonlinear parameterizations.
\subsection{Equivalence theorem and unitarity: an example}
\label{sec:et-example}
As example for the ET in a nonlinear parameterization, we consider top
production by vector boson fusion~\cite{Alcaraz:2000xr}, following the discussion of vector
boson scattering in~\cite{Valencia:1990jp,Seymour:1995qg}.  This example has
also been discussed in~\cite{Veltman:1990} for the
heavy Higgs limit in a linear parameterization,
and an effective field theory
analysis of effects of a non-standard Higgs boson has been given
in~\cite{Han:2000ic}.  According to the flip formalism, in the NL-SM
there are two separately gauge invariant sets of diagrams
\begin{equation}\label{eq:top-pair-groves}
\begin{aligned}
  G_g&=\left\{
\parbox{25\unitlength}{
\fmfframe(5,5)(5,5){
\begin{fmfchar*}(15,15)
\fmfleft{l1,l2}
\fmfright{r1,r2}
\fmf{photon}{l1,fwf1}
\fmf{photon}{fwf2,l2}
\fmf{fermion}{r1,fwf1,fwf2,r2}
\fmfdot{fwf1}
\fmfdot{fwf2}
\fmfv{label=$t $,la.si=right}{r2}
\fmfv{label=$\bar t$,la.si=right}{r1}
\fmfv{label=$W^-$,la.di=0.1cm}{l1}
\fmfv{label=$W^+$,la.di=0.1cm}{l2}
\end{fmfchar*}}}\quad,\quad
  \parbox{30\unitlength}{
\fmfframe(5,5)(5,5){
\begin{fmfgraph*}(20,15)
\fmfleft{f1,f2}
\fmfright{a,b}
\fmf{fermion}{a,i1,b}
\fmf{photon}{f1,i2,f2}
\fmf{photon,label=$Z/\gamma$}{i1,i2}
\fmfdot{i1,i2}
\fmfv{label=$\bar  t$}{a}
\fmfv{label=$t$}{b}
\fmfv{label=$W^-$,la.di=0.1cm}{f1}
\fmfv{label=$W^+$,la.di=0.1cm}{f2}
\end{fmfgraph*}}}
\right\}\\ 
 G_H&=\left\{
  \parbox{30\unitlength}{
\fmfframe(5,5)(5,5){
\begin{fmfgraph*}(20,15)
\fmfleft{f1,f2}
\fmfright{a,b}
\fmf{fermion}{a,i1,b}
\fmf{photon}{f1,i2,f2}
\fmf{dashes,label=$H$}{i1,i2}
\fmfdot{i1,i2}
\fmfv{label=$\bar  t$}{a}
\fmfv{label=$t$}{b}
\fmfv{label=$W^-$,la.di=0.1cm}{f1}
\fmfv{label=$W^+$,la.di=0.1cm}{f2}
\end{fmfgraph*}}}
\right\}
\end{aligned}
\end{equation}
As we will now show, the ET holds separately for both classes of
diagrams.  As a caveat, the additional GICs in the NL-SM do not
necessarily have a good high energy behavior taken by themselves since
in nonlinear parameterizations tree-level unitarity is not a
consequence of gauge invariance.  According
to~\cite{LlewellynSmith:1973}, the only theories of massive vector
bosons respecting tree-level unitarity---i.e. the requirement that the
tree level matrix elements for $N$-particle scattering amplitudes at
high energies scale at most as~$E^{4-N}$---are equivalent to linearly
parameterized spontaneously broken gauge theories.  In the NL-SM with
standard Higgs couplings, reparameterization invariance of the $S$
matrix implies that the \emph{complete} set of diagrams satisfies the
tree-level unitarity bound, so the classes of diagrams that show good
high energy behavior will in general be the same in both
parameterizations.  This will become apparent in the example below.

In the NL-SM, the Higgs-gauge boson and Higgs-GB 
vertices arise from the operator
\begin{multline}\label{eq:nl-phihw}
\frac{1}{2}vH\tr\lbrack (D_\mu U)^\dagger D^\mu U\rbrack\\
= \frac{g}{m_W}H(\partial^\mu\phi^++
m_W W^{+,\mu})(\partial_\mu\phi^-+m_W
W^-_\mu)+\dots
\end{multline}
that includes also a similar term involving the $Z$ and $\phi^0$
bosons and terms of higher order of the GBs not shown here. Here we
have defined $U=\exp\left(\ii\frac{\vec
    \phi\cdot\vec\sigma}{v}\right)$ and the covariant derivative is
given by $D_\mu U=\partial_\mu U+\ii g W^i_\mu\tfrac{\sigma^i}{2}U -\ii
g'UB\tfrac{\sigma^3}{2}$.  It can be checked that the vertices
obtained from~\eqref{eq:nl-phihw} satisfy the WIs~\eqref{eq:h-wis}.
Since the operator~\eqref{eq:nl-phihw} is gauge invariant by itself, in an
effective field theory approach to a nonstandard
Higgs~\cite{Chivukula:1993}, it can be included with an arbitrary
coefficient.  The Yukawa couplings arise from the operators
\begin{multline}\label{eq:nl-yukawa}
\mathcal{L}_Y=-\bar Q_L U\left[\tfrac{(m_t+m_b)}{2}
  +\tfrac{(m_t-m_b)}{2}\sigma^3\right]Q_R\\ -
H\bar Q_L U\left[\tfrac{(\lambda_t+\lambda_b)}{2} +\tfrac{(\lambda_t-\lambda_b)}{2}\sigma^3\right]Q_R+\text{h.c.}
\end{multline}
where gauge invariance allows for Higgs-Yukawa couplings $\lambda_{t,b}$
not related to the fermion masses $m_{t,b}$. Neglecting the bottom mass, 
the relevant interaction terms are given by
\begin{multline}\label{eq:top-yukawa}
\mathcal{L}_Y= - \frac{\ii m_t}{\sqrt 2 v}\phi^+ 
\bar t \left(\tfrac{1-\gamma^5}{2}\right)b
+\frac{\ii m_t}{\sqrt 2 v}\phi^- \bar b \left(\tfrac{1+\gamma^5}{2}\right) t\\
-\lambda_t H\bar t t+ \frac{m_t}{v^2}\phi^+\phi^-\bar t t+\dots
\end{multline}
We will use the  ET to calculate the contributions that grow with the energy
and potentially can violate the unitarity bound, i.e. we 
calculate the diagrams for $\phi^+\phi^-\to t\bar t$ in the limit 
$m_W\to 0$ with $v=\frac{2m_W}{g}=$const. while keeping the top mass fixed.
While the $t$-channel diagram shows no dangerous high energy behavior, the
Higgs exchange diagram and the contact diagram from the interaction quadratic
in the GBs in~\eqref{eq:top-yukawa} grow linearly with the energy\footnote{
  Recall that the the spinors $t$ scale with $\sqrt E$}:
\begin{multline}\label{eq:top-gb}
\parbox{25\unitlength}{
\fmfframe(5,5)(5,5){
\begin{fmfgraph*}(15,15)
\fmfleft{f1,f2}
\fmfright{a,b}
\fmf{fermion}{a,i1,b}
\fmf{dashes}{f2,i1,f1}
\fmfdot{i1}
\fmfv{label=$\bar t$}{a}
\fmfv{label=$ t$}{b}
\fmfv{label=$\phi^-$}{f1}
\fmfv{label=$\phi^+$}{f2}
\end{fmfgraph*}}}
  \,+\,
 \parbox{20\unitlength}{
\fmfframe(2,2)(2,2){
\begin{fmfgraph*}(15,15)
\fmfleft{f1,f2}
\fmfright{a,b}
\fmf{fermion}{a,i1,b}
\fmf{dashes}{i2,f2}
\fmf{dashes}{i2,f1}
\fmf{dashes,label=$H$}{i1,i2}
\fmfdot{i1,i2}
\fmfv{label=$\bar  t$}{a}
\fmfv{label=$t$}{b}
\fmfv{label=$\phi^-$}{f1}
\fmfv{label=$\phi^+$}{f2}
\end{fmfgraph*}}}\\
=\ii 
\left(\frac{m_t}{v^2}\right)\bar t t \left[1-
\left(\frac{\lambda_t v}{m_t}\right)
\frac{s}{s-m_H^2+\ii\Im\Pi_H(s)}\right] 
\end{multline}
where we have used the Dyson summation to introduce the imaginary part
of the Higgs self energy $\Pi_H$ into the propagator.  As can be
checked using the Feynman rules obtained from~\eqref{eq:nl-phihw}, the
Higgs exchange diagram is reproduced by the corresponding diagram with
external longitudinal $W$ bosons in the limit where the gauge boson
mass is sent to zero.  The contact diagram is reproduced from the
$t$-channel diagram and the $s$-channel $Z/\gamma$ exchange diagrams
after a nontrivial cancellation of terms growing like
$E^2$~\cite{Veltman:1990}.  Therefore this diagram is connected to the
grove $G_g$ in~\eqref{eq:top-pair-groves} by the ET.  This shows that,
in agreement with the expectation from the flip formalism, in the
NL-SM the ET holds separately for the groves $G_g$ and $G_H$
in~\eqref{eq:top-pair-groves}, also for a finite width.  However, to
obtain good high energy behavior, all diagrams from both $G_g$ and
$G_H$ have to be considered.  Furthermore, only for the SM value of
the Higgs Yukawa coupling $\lambda_t=\frac{m_t}{v}$ and a vanishing
width both diagrams add up to an amplitude with good high energy
behavior proportional to $\frac{m_H^2}{s-m_H^2}$.  In an effective
field theory approach, unitarity up to the cutoff of the effective
theory implies bounds on the nonstandard couplings~\cite{Han:2000ic}
like $\lambda_t$.

In a linear parameterization, the contact diagram appearing
in~\eqref{eq:top-gb} is absent while in the numerator of the Higgs
exchange diagram $s$ is replaced by $m_H^2$ so for a vanishing width 
the same result for the
GB amplitude is obtained as in the NL-SM \emph{without cancellation
  among diagrams}.  However, the GB amplitude agrees with the one for
the longitudinal gauge bosons only for a vanishing width so the Dyson
summation is incompatible with the naive ET in the linear
parameterization.
 
Since the imaginary part of the gauge boson contributions to the Higgs
self energy is proportional to $s^2$~\cite{Seymour:1995qg}, the
result~\eqref{eq:top-gb} violates the unitarity bound when a realistic
expression for $\Pi_H$ is inserted.  To include a running width
without violating unitarity in the NL-SM, a modified Higgs
propagator~\cite{Seymour:1995qg}
\begin{equation}\label{eq:higgs-rw}
  D_H(q^2)\to \frac{\ii (1+\ii\gamma_H)}{q^2-m_H^2+\ii\gamma_H q^2}
\quad \text{with} \,\gamma_H=\Gamma_H/M_H \theta(q^2)
\end{equation}
has been proposed in the context of gauge boson scattering (for a
generalization beyond leading order see~\cite{Kilian:1998bh}).  For
the SM value of $\lambda_t$, using this form of the propagator allows
the cancellation between the Higgs exchange diagram and the contact
diagram to take place in~\eqref{eq:top-gb} and one obtains an
amplitude proportional to $\frac{m_H^2}{s(1+\ii\gamma_H)-m_H^2}$.
Therefore good high energy behavior is restored also for a running
width.  This has to be compared to the case of the $W$ boson, where
the introduction of a running width~\cite{Baur:1995} similar
to~\eqref{eq:higgs-rw} is in general incompatible with gauge invariance unless
all radiative corrections are included in the same order,  resulting
in possibly large numerical errors in certain regions of phase 
space~\cite{Argyres:1995}.  In a linear
parameterization, also the propagator~\eqref{eq:higgs-rw} is not
compatible with the naive ET.

To summarize, in a nonlinear parameterization there are additional
GICs compared to a linear one that satisfy the ET by themselves, also
for a finite width of the Higgs boson.  However, the classes of
diagrams that are gauge invariant \emph{and} show good high energy
behavior will in general be the same in both parameterizations.
 
\subsection{Compatibility of the ET with Dyson summation: general discussion}
\label{sec:et-proof}
After this example, we proceed to a general proof of the consistency
of the Dyson summation of the Higgs propagator in the NL-SM
and discuss the relation to the conjectured simplification of the 
ET~\cite{Valencia:1990jp} in more detail.  To be precise,
in~\cite{Valencia:1990jp} it has been suggested that matrix elements
can contain external Higgs bosons off their mass shell without
violating the ET, provided a nonlinear parameterization is used.  This
property is in fact a straightforward consequence of the trivial BRST
transformation $\delta_{\text{BRST}} H=0$ of the Higgs in the NL-SM.  
Recall that the derivation
of the ET requires the WIs for amplitudes with insertions of the
operator $(\ii p_\mu
V_a^\mu+m_{V_a}\phi_a)$~\cite{Chanowitz:1985}.  In the BRST
formalism, they are derived using the Kugo-Ojima condition that the
BRST charge $Q$ annihilates physical states.  This results in STIs of
the form 
\begin{equation}
0=\braket{\text{out}|\Tprod{\{Q,\bar c_a B_b\dots
    B_n\}}|\text{in}}=\braket{\text{out}|\Tprod{B_a\dots B_n}|\text{in}}
\end{equation}
 where $B_a$ is
the Nakanishi-Lautrup auxiliary field obtained from the BRST
transformation of an antighost $\bar c_a$.  We use a linear $R_\xi$
gauge fixing also in the nonlinear parameterization, so that the
equation of motion of $B_a$ is given by $B_a=-\frac{1}{\xi}(\partial_\mu V_a-\xi
m_{V_a}\phi_a)$ as in the linear parameterization.  In the NL-SM, the
trivial BRST transformation of the Higgs implies that similar
identities are true also if additional Higgs boson field operators are
inserted in the Green's functions:
\begin{multline}\label{eq:wi-et}
0=\braket{\text{out}|\Tprod{\{Q,\bar c_a B_b\dots B_n H\dots H\}}|\text{in}}\\
=\braket{\text{out}|\Tprod{B_a\dots B_n H\dots H}|\text{in}}
\end{multline}
In a linear parameterization,  the
BRST transformation $\delta_{\text{BRST}} H=\frac{g}{2}(c^+\phi^-+c^-
\phi^+)+\frac{g}{2\cos\theta_w}c_Z\phi_0$ mixes the Higgs with the GBs
so there are additional terms
on the right hand side of~\eqref{eq:wi-et}.

The WIs~\eqref{eq:wi-et} allow in the usual
way~\cite{Chanowitz:1985,Dobado:1994} to deduce the validity of the
ET also for matrix elements with external off-shell Higgs bosons
$H^\star$, in agreement with the conjecture of~\cite{Valencia:1990jp}:
\begin{multline}\label{eq:higgs-et}
  \mathcal{M}(\text{in} \to
  \text{out}+V^L_a\dots V^L_n H^*\dots H^* )\\
= (-\ii)^n\mathcal{M}(\text{in}\to
  \text{out}+\phi_a\dots\phi_n H^*\dots H^*  )\\
+\mathcal{O}\left(\frac{m_V}{E}-\text{suppressed}\right)
\end{multline}
where one phase of $(-\ii)$ occurs for every outgoing longitudinal
gauge boson $V^L$ (for incoming gauge bosons, the sign of the phase
has to be changed) and we have suppressed renormalization
factors~\cite{Yao:1988}.  While in a renormalizable theory and for
external on-shell Higgs bosons, the
additional contributions on the right hand side are of order
$O\left(\frac{m_V}{E}\right)$ and the GB amplitude is bounded at
large energies, in a nonlinearly parameterized effective field theory,
the additional contributions are suppressed only relative to the GB
amplitude~\cite{Dobado:1994} that need not show good high energy behaviour.
To assess the usefulness of the ET in a given situation, the absolute
value of the additional terms has to be estimated for the process
and the energy range of interest~\cite{Dobado:1994}.
We always take it as understood that the ET in a nonlinearly parameterized
theory holds in this restricted sense.

 We now demonstrate that~\eqref{eq:higgs-et} already implies also the
 validity of the ET for internal off-shell Higgs bosons, also when a
 Dyson resummed propagator is used.  To show this, we give a
 diagrammatic prescription to express the complete amplitude in terms
 of matrix elements with off shell Higgs bosons and all other
 particles on the mass shell. Consider the set of diagrams that has in
 common an internal Higgs boson line (that is not part of a closed
 loop) with a given momentum $p_H$, labeling the external momenta such
 that $-(p_1+\dots+p_i)=p_H=p_{i+1}+\dots+ p_n$ and treating all
 momenta as incoming.  This set of diagrams can be written in the
 factorized form
\begin{multline}\label{eq:higgs-factorize}
\parbox{45\unitlength}{
\fmfframe(5,5)(5,5){
\begin{fmfgraph*}(35,17)
\fmfleftn{l}{3}
\fmfrightn{r}{3}
\begin{fmffor}{i}{1}{1}{3}
\fmf{plain}{l[i],i1}
\fmf{plain}{r[i],i2}
\end{fmffor}
\fmf{dashes,tension=2}{i1,b}
\fmf{dashes,tension=2}{b,i2}
\fmfv{d.sh=circle,d.f=empty,d.size=20pt}{i1}
\fmfv{d.sh=circle,d.f=empty,d.size=20pt}{i2}
\fmfv{d.sh=circle,d.f=shaded,d.size=10pt,la=$p_H$,la.an=90,la.di=10pt}{b}
\fmfv{la=$p_i$}{l1}
\fmfv{la=$\vdots$,la.an=160}{l2}
\fmfv{la=$p_1$}{l3}
\fmfv{la=$p_{n}$}{r1}
\fmfv{la=$\vdots$,la.an=10}{r2}
\fmfv{la=$p_{i+1}$}{r3}
\end{fmfgraph*}}}\\
= \,\mathcal{M}(\Psi_1\dots\Psi_i H^*)\,D_H(p_H)\, 
  \mathcal{M}(H^*\Psi_{i+1}\dots\Psi_n)
\end{multline}
We have depicted the case of a $s$-channel Higgs boson, but a similar
decomposition holds for $t$-or $u$-channel Higgs lines.  If the
complete amplitude is evaluated at a given loop order, the
decomposition~\eqref{eq:higgs-factorize} has to be understood as
consistently expanded up to this order. 
Iterating the decomposition~\eqref{eq:higgs-factorize} by applying the
same formula to the sub-amplitudes\footnote{There is a subtlety in
  avoiding double counting of diagrams. For instance, first
  applying~\eqref{eq:higgs-factorize} to a Higgs boson line with some
  momentum $p_{H_1}$ and subsequently factorizing another momentum
  $p_{H_2}$ out of a sub-amplitude yields contributions that appear
  also when the contribution of $p_{H_2}$ is factorized first. Such
  contributions generated more then once have to be omitted} we arrive
at a decomposition of the amplitude in terms of matrix elements with
off-shell Higgs bosons and all other external particles on the
mass-shell. When applied to arbitrary internal particles off the mass
shell, the individual terms of such a decomposition are in general not
gauge invariant by themselves.  For the case of Higgs bosons in the
NL-SM, however, the sub-amplitudes occurring
in~\eqref{eq:higgs-factorize} are precisely the quantities satisfying
the ET for off shell Higgs bosons~\eqref{eq:higgs-et}.  As a
consequence, the contributions to the $S$-matrix with internal Higgs
boson lines with a given set of momenta satisfy the ET by themselves.
Since this property is
independent of the expression used for the Higgs propagators $D_H$
in~\eqref{eq:higgs-factorize}, we can use the Dyson resummed
propagator or a simple effective prescription
like~\eqref{eq:higgs-rw} without violating WIs or the ET.
As in the example of subsection~\ref{sec:et-example}, the subsets of diagrams
with internal Higgs bosons need not respect unitarity bounds. 
Nevertheless, the separate gauge invariance can be useful for the discussion
of simple schmes to restore unitarity.

In addition to the simple WIs~\eqref{eq:wi-et}, the trivial BRST
transformation law of the Higgs boson implies the gauge parameter
independence of Green's functions with off-shell Higgs bosons, if all
other external particles are on-shell.  As reviewed briefly in
appendix~\eqref{app:nielsen}, in the BRST formalism the gauge
parameter dependence of Green's functions can be expressed in terms of
Green's functions with insertions of BRST transformed
fields~\eqref{eq:nielsen-gf}.  From this, one obtains immediately
\begin{equation}\label{eq:higgs-gpi}
  \partial_\xi\braket{\text{out}|\Tprod{H\dots H}|\text{in}}=0
\end{equation}
By the same reasoning as above, this implies gauge parameter
independence of subsets of diagrams with a given set of momenta of
internal Higgs lines, independent of the prescription used for the
Higgs propagator (provided it is gauge parameter independent by
itself).  Incidently, these results give an independent derivation of
the classification of GICs in the NL-SM in terms of the number of
internal Higgs boson lines derived in~\cite{OS:FGH} using the flip
formalism.
\section{Properties of the Higgs self energy in the nonlinear parameterization}\label{sec:nl}
Apart from the consistency of the Dyson summation with the ET, 
another result  motivated by the flip formalism in section~\ref{sec:groves}
is the gauge parameter 
independence of the Higgs propagator in the NL-SM:
\begin{equation}\label{eq:higgs-prop-gpi}
  \partial_\xi\Greensfunc{H(x)H(y)}=0 
\end{equation}
In fact, this is merely a special case of the 
gauge parameter independence of matrix elements with off-shell
Higgs bosons~\eqref{eq:higgs-gpi}.
In a linear parameterization, eq.~\eqref{eq:higgs-prop-gpi} 
is violated because the
BRST transformation of the Higgs field is nontrivial 
(see~\eqref{eq:nielsen-gf} and the remarks below~\eqref{eq:wi-et}).
In the presence of CP violating mixing with the gauge boson 
sector, the full Higgs propagator includes contributions from the $Z$ and
GB propagators~\cite{Pilaftsis:1997dr}:
\begin{multline}
  \label{eq:higgs-dyson}
\Greensfunc{H(x)H(y)}
=\,
\parbox{15mm}{
\begin{fmfgraph*}(15,15)
\fmfleft{l}
\fmfright{r}
\fmf{dashes}{l,i,r}
\fmfv{d.sh=circle,d.f=shaded,d.size=10pt,la=$\Gamma_{HH}$,la.ang=-75,la.di=0.5cm}{i}
\end{fmfgraph*}}
\,+\,
\parbox{25mm}{
\begin{fmfgraph*}(25,15)
\fmfleft{l}
\fmfright{r}
\fmf{dashes,tension=1.5}{l,i1}
\fmf{dashes,label=$H$,la.si=left}{i1,i2}
\fmf{dashes,tension=1.5}{i2,r}
\fmfv{d.sh=circle,d.f=shaded,d.size=10pt,la=$\Gamma_{HH}$,la.ang=-75,la.di=0.5cm}{i1}
\fmfv{d.sh=circle,d.f=shaded,d.size=10pt,la=$\Gamma_{HH}$,la.ang=-75,la.di=0.5cm}{i2}
\end{fmfgraph*}}\\
+\,
\parbox{25mm}{
\begin{fmfgraph*}(25,15)
\fmfleft{l}
\fmfright{r}
\fmf{dashes,tension=1.5}{l,i1}
\fmf{photon,label=$Z$,la.si=left}{i1,i2}
\fmf{dashes,tension=1.5}{i2,r}
\fmfv{d.sh=circle,d.f=shaded,d.size=10pt,la=$\Gamma_{HZ}$,la.ang=-75,la.di=0.5cm}{i1}
\fmfv{d.sh=circle,d.f=shaded,d.size=10pt,la=$\Gamma_{ZH}$,la.ang=-75,la.di=0.5cm}{i2}
\end{fmfgraph*}}
\,+\,
\parbox{25mm}{
\begin{fmfgraph*}(25,15)
\fmfleft{l}
\fmfright{r}
\fmf{dashes,tension=1.5}{l,i1}
\fmf{dashes,label=$\phi^0$,la.si=left}{i1,i2}
\fmf{dashes,tension=1.5}{i2,r}
\fmfv{d.sh=circle,d.f=shaded,d.size=10pt,la=$\Gamma_{H\phi^0}$,la.ang=-75,la.di=0.5cm}{i1}
\fmfv{d.sh=circle,d.f=shaded,d.size=10pt,la=$\Gamma_{\phi^0 H}$,la.ang=-75,la.di=0.5cm}{i2}
\end{fmfgraph*}}\,+\,\dots
\end{multline}
Therefore the gauge parameter
independence of the full propagator~\eqref{eq:higgs-prop-gpi}
will in general be a consequence of cancellations among the different
contributions and the Higgs self energy by itself can be gauge
parameter dependent.
 This will be discussed from the perspective of
the flip formalism below.

But first, let us demonstrate explicitly the cancellation of the gauge
parameter in the one loop gauge boson contribution to the Higgs propagator.
We decompose the gauge boson propagator in $R_\xi$ gauge into the
propagator in unitarity gauge and a term
proportional to the GB propagator, introducing the
graphical notation
\begin{equation}
\label{eq:uni-xi-graph}
\begin{aligned}
\parbox{10\unitlength}{
\begin{fmfgraph}(10,10)
\fmfleft{l}
\fmfright{r}
\fmf{photon}{l,r}  
\end{fmfgraph}}
\quad&=\quad
\parbox{15\unitlength}{
\begin{fmfgraph}(15,10)
\fmfleft{l}
\fmfright{r}
\fmf{gluon}{l,r}  
\end{fmfgraph}}
&+&\quad\parbox{15\unitlength}{
\begin{fmfgraph}(15,10)
\fmfleft{l}
\fmfright{r}
\fmf{dbl_dots}{l,r}  
\end{fmfgraph}}\\
D_{W,\xi}^{\mu\nu}(q)&= \frac{-\ii\left(g^{\mu\nu}-\tfrac{q^\mu
      q^\nu}{m_W^2}\right)}{q^2-m_W^2} &+&\,\left(-\frac{q^\mu
    q^\nu}{m_W^2}\right) \frac{\ii}{q^2-\xi m_W^2}
\end{aligned}
\end{equation}
Because of the trivial BRST transformation, there are no ghost-Higgs
vertices so we only have to consider the gauge boson and GB loops. 
The peculiar form of 
the interaction lagrangian~\eqref{eq:nl-phihw} ensures that 
 the contributions with one unphysical pole add up to zero:
\begin{subequations}\label{eq:one-xi}
\begin{align}
\parbox{15mm}{
\begin{fmfgraph*}(15,15)
\fmfleft{l}
\fmfright{r}
\fmftop{t}
\fmf{dashes,tension=3,label=$p$}{l,v1}
\fmf{dashes,tension=3}{v2,r}
\fmf{dbl_dots,right=1,label=$p+k$}{v2,v1}
\fmf{gluon,right=1,tension=1.5,label=$k$,la.di=0.5cm}{v1,v2}
\fmfdot{v1,v2}
\end{fmfgraph*}}
=&\left(\ii g m_W\right)^2 \int \dd^4 k\,
\frac{-\ii\left(g^{\mu\nu}-\tfrac{k^\mu k^\nu}{m_W^2}\right)}{k^2-m_W^2}
\nonumber\\
&\left(-\ii\frac{(p+k)_\mu(p+k)_\nu}{m_W^2((p+k)^2-\xi m_W^2)}\right)\\
\parbox{15mm}{
\begin{fmfgraph*}(15,15)
\fmfleft{l}
\fmfright{r}
\fmftop{t}
\fmf{dashes,tension=3,label=$p$}{l,v1}
\fmf{dashes,tension=3}{v2,r}
\fmf{dashes,right=1,label=$p+k$,la.di=0.1cm}{v2,v1}
\fmf{gluon,right=1,tension=1.5,label=$k$,la.di=0.5cm}{v1,v2}
\fmfdot{v1,v2}
\end{fmfgraph*}}=&g^2\int \dd^4 k\, (p+k)_\mu 
\frac{-\ii\left(g^{\mu\nu}-\tfrac{k^\mu k^\nu}{m_W^2}\right)}{k^2-m_W^2}\\\nonumber
&\frac{\ii}{(p+k)^2-\xi m_W^2}\left(-(p+k)_\nu\right)
\end{align}
\end{subequations}
In the second diagram, one minus sign arises because the gauge boson momentum
is incoming at one vertex and outgoing at the other.
One can show similarly, that the contributions with two unphysical poles
add up to zero.
In contrast, in the linear parameterization the $H\Phi W$ vertex
has the form $\frac{g}{2}W^{\pm,\mu}H\overleftrightarrow{\partial_\mu}\phi^\mp$
and the cancellation doesn't go through as in~\eqref{eq:one-xi}
 so the gauge boson contribution is 
gauge parameter dependent~\cite{Papavassiliou:1998pb}. 

In the presence of Higgs-$Z$ mixing, gauge parameter independence of
the Higgs propagator doesn't already imply gauge parameter
independence of the Higgs self energy to all orders.  As shown in
appendix~\ref{app:nielsen} using the formalism
of~\cite{Piguet:1985,Gambino:1999}, in the presence of
CP violation the gauge parameter dependence of the Higgs self energy
takes the form
\begin{equation}\label{eq:higgs-nielsen}
  \partial_\xi \Gamma_{HH}=2\Lambda_{\phi^{0}H}\Gamma_{\phi^0H}+2\Lambda_{ZH,\mu}\Gamma_{ZH}^\mu
\end{equation}
where $\Lambda_{\phi^{0}H}$ and $\Lambda_{ZH}$ are
vertex functions with insertions of the gauge parameter dependent part of the
gauge fixing functional (see appendix~\ref{app:nielsen} for the precise
expressions). 
In the SM this mixing is phenomenologically not important since it is 
induced by CP violating effects only at the three loop 
level, but the phenomenon will persist in CP violating extensions of the SM.
The discussion is also relevant for the CP odd scalar $A$ in the 2HDM
discussed in section~\ref{sec:2hdm}.
The reason why CP violating Higgs-$Z$ mixing introduces additional
complications can also be understood directly on the level of
gauge flips. 
Once $ZH$ mixing is generated radiatively as in the minimal SM by
an insertion of a box diagram~\cite{Pilaftsis:1997dr}, 
there is a gauge flip to 
vertex correction diagrams:
\begin{multline}
\parbox{40mm}{
\begin{fmfgraph*}(40,20)
\fmfright{r1,r,r2}
\fmfleft{l}
\fmftop{t1,t2}
\fmfbottom{b1,b2}
\fmf{fermion,right=0.5,tension=0.5,label=$c$,la.si=right}{i1,i2}
\fmf{fermion,tension=5,label=$b$,la.si=right}{i2,i3}
\fmf{fermion,right=0.5,tension=1,label=$t$,la.si=right}{i3,i4,i5}
\fmf{fermion,tension=5,label=$s$,la.si=right}{i5,i6}
\fmf{fermion,right=0.5,tension=0.5,label=$c$,la.si=right}{i6,i1}
\fmf{dashes,tension=2,label=$H$}{i4,l}
\fmf{phantom,tension=0.5}{r,i1}
\fmf{phantom,tension=2}{t1,i3}
\fmf{phantom,tension=1.5}{t2,i2}
\fmf{phantom,tension=2}{b1,i5}
\fmf{phantom,tension=1.5}{b2,i6}
\fmffreeze
\fmf{plain}{r1,v,r2}
\fmf{photon,label=$Z$}{v,i1}
\fmf{photon,label=$\scriptstyle{W^-}$,la.si=right,la.di=0.07cm}{i3,i5}
\fmf{photon,label=$\scriptstyle{W^+}$,la.si=left,la.di=0.07cm}{i2,i6}
\fmfdot{v,i1,i2,i3,i4,i5,i6}
\end{fmfgraph*}}\Leftrightarrow\,
\parbox{40mm}{
\begin{fmfgraph*}(40,20)
\fmfright{r1,r,r2}
\fmfleft{l}
\fmftop{t1,t2}
\fmfbottom{b1,b2}
\fmf{photon,tension=0.5,label=$\scriptstyle{W^+}$,la.si=right,la.di=0.1cm}{i1,i2}
\fmf{fermion,tension=5,label=$b$,la.si=right}{i2,i3}
\fmf{fermion,right=0.5,tension=1,label=$t$}{i3,i4,i5}
\fmf{fermion,tension=5,label=$s$}{i5,i6}
\fmf{photon,tension=0.5,label=$\scriptstyle{W^+}$,la.si=right}{i6,i1}
\fmf{dashes,tension=2,label=$H$}{i4,l}
\fmf{phantom,tension=0.5}{r,i1}
\fmf{phantom,tension=2}{t1,i3}
\fmf{phantom,tension=1.5}{t2,i2}
\fmf{phantom,tension=2}{b1,i5}
\fmf{phantom,tension=1.5}{b2,i6}
\fmffreeze
\fmf{plain}{r1,v,r2}
\fmf{photon,label=$Z$}{v,i1}
\fmf{photon,label=$\scriptstyle{W^-}$,la.si=right,la.di=0.05cm}{i3,i5}
\fmf{fermion,label=$c$,la.si=right}{i2,i6}
\fmfdot{v,i1,i2,i3,i4,i5,i6}
\end{fmfgraph*}}\\
\Leftrightarrow\,
\parbox{40mm}{
\begin{fmfgraph*}(40,20)
\fmfright{r1,r2}
\fmfleft{l}
\fmftop{t1,t2}
\fmfbottom{b1,b2}
\fmf{photon,tension=1,label=$\scriptstyle{W^+}$,la.si=right}{v2,i2}
\fmf{fermion,tension=5,label=$b$,la.si=right}{i2,i3}
\fmf{fermion,right=0.5,tension=1,label=$t$}{i3,i4,i5}
\fmf{fermion,tension=5,label=$s$}{i5,i6}
\fmf{photon,tension=1,label=$\scriptstyle{W^+}$,la.si=right}{i6,v1}
\fmf{dashes,tension=2,label=$H$}{i4,l}
\fmf{phantom,tension=2}{t1,i3}
\fmf{phantom,tension=1.5}{t2,i2}
\fmf{phantom,tension=2}{b1,i5}
\fmf{phantom,tension=1.5}{b2,i6}
\fmf{plain,tension=2}{r1,v1,v2,r2}
\fmffreeze
\fmf{photon,label=$\scriptstyle{W^-}$,la.si=right,la.di=0.05cm}{i3,i5}
\fmf{fermion,label=$c$,la.si=right}{i2,i6}
\fmfdot{v1,v2,i2,i3,i4,i5,i6}
\end{fmfgraph*}}
\end{multline}
Since the full resummed Higgs propagator~\eqref{eq:higgs-dyson}
includes also contributions from Higgs-$Z$ mixing, similar flips
connect the irreducible Higgs self energy to reducible diagrams with
Higgs-$Z$ mixing~(compare to~\eqref{eq:selfenergy-flip}) so once CP violation
occurs, the Higgs self energy is not expected to be gauge parameter
independent by itself.  Nevertheless, the gauge parameter dependence
must cancel between the different irreducible two point functions
since the full Higgs propagator is gauge parameter independent.  

To clarify this issue further, one can demonstrate the cancellation among the
various contributions by applying the formalism
of~\cite{Piguet:1985,Gambino:1999} to the full
propagator~\eqref{eq:higgs-dyson} including mixing.  The complete treatment to
all orders involves a three by three matrix describing the mixing among
$Z$,$\phi^0$ and $H$ and is beyond the scope of this note.  Here we give a
simplified analysis valid in the first loop order $n$ where $HZ$ mixing is
non vanishing. In this case we can restrict ourselves to the diagrams shown
in~\eqref{eq:higgs-dyson} and need not consider the mixing of the $Z$ boson
with the GB $\phi^0$.
The variation of 
the Higgs propagator with respect to the gauge parameter receives
contributions from the gauge boson and GB propagators and from the
self-energies themselves:
\begin{multline}
  \partial_\xi D_H^{(2n)} = D_H^{(0)}\Bigl[ \partial_\xi\Gamma_{HH}^{(2n)}
    +2\left(\partial_\xi\Gamma_{HZ}^{(n),\mu}\right)
    D_{Z,\mu\nu}^{(0)}\Gamma_{ZH}^{(n),\nu}\\
    +2\left(\partial_\xi\Gamma_{H\phi^0}^{(n)}\right)D_{\phi^0}^{(0)}
    \Gamma_{\phi H}^{(n)}\Bigr] D_H^{(0)} \\
  +D_H^{(0}\left[\Gamma_{HZ}^{(n),\mu} (\partial_\xi D_{Z,\mu\nu}^{(0)}
    )\Gamma_{ZH}^{(n),\nu}
    +\Gamma_{H\phi^0}^{(n)}(\partial_\xi D_{\phi^0}^{(0)})
    \Gamma_{\phi^0 H}^{(n)}\right] D_H^{(0)}
\end{multline}
Here $D_{H}^{(2n)}$ denotes the Higgs propagator up to order $2n$, where $n$
is first order where $\Gamma_{HZ}$ is non vanishing.  In the order considered,
the propagators of the $Z$ and $\phi^0$ are tree level propagators so we can
use the explicit expression~\eqref{eq:uni-xi-graph} to verify that the terms
involving the variation of the propagators cancel among themselves because of
the simple WI $(\ii p_\mu \Gamma_{ZH}^\mu+m_Z\Gamma_{\phi^0H})=0$.  To
simplify the remaining terms we use the Identity~\eqref{eq:neutral-nielsen},
making the plausible assumption that the vertex functions $\Lambda_{HZ}$ and
$\Lambda_{H\phi^0}$ arise only at the same order as the mixing $\Gamma_{ZH}$.
Up to this order the two point vertex functions enter only on tree level and
we obtain
\begin{equation}
    \partial_\xi \Gamma_{HZ}^{(n),\mu}=
\Lambda_{HZ,\nu}^{(n)}\Gamma_{ZZ}^{(0),\nu\mu}
   \quad \partial_\xi \Gamma_{H\phi^0}^{(n)}=
\Lambda_{H\phi^0}^{(n)}\Gamma_{\phi^0\phi^0}^{(0)}
\end{equation}
Since the two point vertex functions are the negative of the inverse
propagators, the gauge parameter dependence of the mixing contributions
cancels against the variation of the Higgs self
energy~\eqref{eq:higgs-nielsen}:
\begin{multline}
 \partial_\xi D_H^{(2n)} = D_H^{(0)}\Bigl[
  \partial_\xi\Gamma_{HH}^{(2n)} \\
  -2\left(\Lambda_{HZ,\mu}^{(n)}
  \Gamma_{ZH}^{(n),\mu} +\Lambda_{H\phi^0}^{(n)}
  \Gamma_{\phi H}^{(n)}\right)\Bigr] D_H^{(0)}=0
\end{multline}
Therefore the gauge parameter dependence of the full propagator vanishes, in
agreement with~\eqref{eq:higgs-prop-gpi}.  For this, it is necessary to
calculate the Higgs-self energy up to the order $2n$, a residual gauge
dependence remains if it is evaluated at the same order as the $HZ$ mixing.
\section{Two-Higgs doublet models}\label{sec:2hdm}
As an important example for a non-minimal Higgs sector, 
in this section we 
discuss to which extent the results obtained for the NL-SM
carry over to a two-Higgs doublet model.
A complication compared to the NL-SM is the appearance of  
vertices involving two Higgs bosons and a gauge boson 
like $W^{+,\mu} (H^0\overleftrightarrow{\partial_\mu} H^-)$
that can lead to a more complicated structure
of GICs~\cite{OS:FGH}.
As we will demonstrate below, in the 2HDM only
the neutral Higgs bosons can be treated as in the 
NL-SM and their Dyson summation doesn't violate gauge invariance.
Our discussion is mainly phrased
in the language of gauge flips, but the
formal proofs of sections~\ref{sec:et} and~\ref{sec:nl} can easily be extended
to the case of the neutral Higgs bosons in the 2HDM since the main 
ingredient is the trivial 
BRST transformation of the neutral Higgs. We also give an explicit example
for the violation of the naive ET by the introduction of a finite
width for the charged Higgs boson.
\subsection{Nonlinear parameterization of the 2HDM}
We will briefly review the 2HDM in the nonlinear parameterization 
introduced in~\cite{Ciafaloni:1997ur} and determine the BRST
transformations of the Higgs bosons that are used in the subsequent
subsections to derive the STIs and the form of the gauge flips.
Following~\cite{Ciafaloni:1997ur} we collect both scalar doublets
$H_1,H_2$ of the 2HDM in a matrix
\begin{equation}
\begin{aligned}
\Phi=(\tilde H_2H_1) \quad&\text{with} \quad \tilde H_i=\ii\sigma_2H_i^* \\
&\text{and} \quad\braket{H_{1,2}}=
\begin{pmatrix}
  0\\ v_{1,2}
\end{pmatrix}
\end{aligned}
\end{equation}
and introduce the nonlinear parameterization $\Phi=U\mathcal{H}$ 
where the GB matrix is again $U=\exp\left(\ii\frac{\vec
    \phi\cdot\vec\sigma}{v}\right)$ and the Higgs bosons are collected
in the matrix
\begin{equation}
\begin{aligned}
&\mathcal{H}=\mathcal{H}_0 +
(h+\ii A^0+\vec\sigma\cdot\vec H)
\begin{pmatrix}\cos\beta&0 \\ 0&\sin\beta\end{pmatrix}\\
 &\text{with}\quad \mathcal{H}_0=\begin{pmatrix} v_2&0 \\ 0&v_1\end{pmatrix}
\end{aligned}
\end{equation}
Here the mixing angle $\tan\beta=\frac{v_2}{v_1}$ has been introduced 
and we define $v$ so that
$v_1=v\cos\beta$ and $v_2=v\sin\beta$.
The mass eigenstates of the neutral scalars
are linear combinations of  $h$ and $H^3$
but the precise form~\cite{Ciafaloni:1997ur} is not needed for our discussion.
More important for us are the `interaction 
eigenstates'~\cite{Ciafaloni:1997ur} $H$ and $S$ defined by
\begin{equation}\label{eq:i-eigen}
  \begin{aligned}
 \cos\beta S+\sin\beta H&=\cos\beta \left(\tfrac{h+H^3}{2}\right)\\  
 -\sin\beta S+\cos\beta H&=\sin\beta \left(\tfrac{h-H^3}{2}\right)
  \end{aligned}
\end{equation}
that will simplify the Feynman rules.
The GB and Higgs matrices  transform  under  $SU(2)_L\times U(1)_Y$ as
\begin{equation}\label{eq:nl-trans}
\Phi\to L\Phi R^\dagger\qquad \quad U\to LUR^\dagger\qquad
\mathcal{H}\to R\mathcal{H}R^\dagger 
\end{equation}
with $L=e^{\ii \vec\alpha\cdot\frac{\vec\sigma}{2}}\in SU(2)_L$ and $
R=e^{\ii\beta\frac{\sigma^3}{2}}\in U(1)_Y$.
Therefore the BRST 
transformation of the Higgs bosons is given by: 
\begin{equation}
\delta_{\text{BRST}}\mathcal{H}=\ii g'c^3\left\lbrack\frac{\sigma^3}{2},\mathcal{H}\right\rbrack
\end{equation}
so the BRST transformations of the charged Higgs bosons 
$H^\pm=\frac{1}{\sqrt 2}(H^1\mp \ii H^2)$ are found to be
\begin{equation}\label{eq:higgs-brst}
\delta_{\text{BRST}} H^\pm=\pm\ii(ec_A-\tfrac{g}{\cos\theta_w}\sin^2\theta_wc_Z) H^\pm
\end{equation}
while the neutral Higgs bosons transform trivially.
Below, we need the  Feynman rules of the Higgs bosons
appearing in the decomposition of the kinetic term
\begin{equation}
 \mathcal{L}_{\text{kin}}=\frac{1}{4}\tr\left[(D_\mu\Phi)^\dagger D^\mu \Phi\right]:=\mathcal{L}_{\mathcal{H}}+\mathcal{L}_{U}
+\mathcal{L}_{\mathcal{H}U}
\end{equation}
where we have defined the operators
\begin{subequations}\label{eq_2hdm-kinetic}
\begin{align}
\mathcal{L}_{\mathcal{H}}&= \frac{1}{4}\tr[D_\mu\mathcal{H}^\dagger D^\mu
 \mathcal{H}]\\
 \mathcal{L}_{U}&=\frac{1}{4}\tr[(D^\mu U^\dagger)(D_\mu U)\mathcal{H}^\dagger\mathcal{H}] \\
 \mathcal{L}_{\mathcal{H}U}&=\frac{1}{4}\left(\tr\left[(U^\dagger D^\mu U)
(\mathcal{H}D_\mu\mathcal{H}^\dagger)\right]+\text{ h.c.}\right)
\end{align}
\end{subequations}
Here the transformation law~\eqref{eq:nl-trans} implies the action of
the covariant derivative on $\mathcal{H}$ as $D_\mu\mathcal{H} =\partial_\mu
\mathcal{H}-\ii g'B_\mu\left\lbrack \frac{\sigma^3}{2}, \mathcal{H}
\right\rbrack$.  Note that the operators in~\eqref{eq_2hdm-kinetic}
are gauge invariant by themselves.  Therefore, in an effective field
theory approach to the 2HDM, the operator $\mathcal{L}_{\mathcal{H}U}$
can appear with an arbitrary coefficient $\lambda_{\mathcal{H}U}$
while the coefficients of the other two operators are fixed by the
normalization of the kinetic terms\footnote{For simplicity,
 here we don't consider the  
introduction of another kinetic term for the GBs of the form $\tr[(D_\mu U)^\dagger D^\mu U]$ that would allow a nonstandard coefficient of $\mathcal{L}_U$.}.
The cubic interaction terms
obtained from the expansion of~\eqref{eq_2hdm-kinetic} will be written
as
\begin{equation}
  \mathcal{L}_{\Phi^3}=\sum_{\Phi_i\Phi_j\Phi_k}
\left(\mathcal{O}_{\Phi_i\Phi_j\Phi_k}^{\mathcal{H}}+\mathcal{O}_{\Phi_i\Phi_j\Phi_k}^{U}+\lambda_{\mathcal{H}U}\mathcal{O}_{\Phi_i\Phi_j\Phi_k}^{\mathcal{H}U}\right)
\end{equation}
where the fields $\Phi$ include the gauge bosons, Higgs bosons and GBs.
Here the operators $\mathcal{O}_{\Phi_i\Phi_j\Phi_k}^{\mathcal{H}}$
arise from the expansion of $\mathcal{L_H}$ and analogously for
$\mathcal{O}_{\Phi_i\Phi_j\Phi_k}^{U}$ and
$\mathcal{O}_{\Phi_i\Phi_j\Phi_k}^{\mathcal{H}U} $.

\subsection{Neutral Higgs bosons}
We now discuss the form of the gauge flips involving neutral Higgs
bosons and the consequences for the GICs and the Dyson summation of
the Higgs propagators.  The charged Higgs bosons are discussed in
subsection~\ref{sec:charged}.  The discussion of the neutral scalars
is simpler in terms of the two interaction eigenstates defined
in~\eqref{eq:i-eigen}. As can be verified from the Feynman rules
arising from $\mathcal{L}_{U}$, the
interaction eigenstate $H$ has the same
interactions with the gauge bosons as the Higgs in the
NL-SM~\eqref{eq:nl-phihw}, while all other scalars have no
interactions of the form $ HV_\mu V^\mu$~\cite{Ciafaloni:1997ur}.  In
the language of the flip formalism, this implies that the flips of the
form~\eqref{eq:higgs-gauge-flip} have to be chosen just like in the
NL-SM and the internal Higgs boson $H$ can be omitted.

We now turn to the additional $HHV$ vertices in the 2HDM.  One finds
that the only vertices of this kind involving neutral Higgs bosons are
contained in the operator $\mathcal{L}_{\mathcal{H}U}$ defined
in~\eqref{eq_2hdm-kinetic} and involve only the second interaction
eigenstate $S$ and the CP odd scalar $A^0$:
\begin{equation}\label{eq:wphmh-vert}
\begin{aligned}
\mathcal{O}^{\mathcal{H}U}_{W^+ H^- H^0}
&:= \frac{1}{\sqrt 2v}(\partial^\mu\phi_++m_W W^{+,\mu})\\
&\Bigl[\frac{1}{2}(A^0\overleftrightarrow{\partial_\mu} H^-)+\ii(S\overleftrightarrow{\partial_\mu} H^-)\Bigr]+\text{ h.c.}\\
\mathcal{O}^{\mathcal{H}U}_{Z S A^0}
&:=\frac{1}{v}
(\partial^\mu\phi_0+m_Z Z^\mu)(S\overleftrightarrow{\partial_\mu} A^0)
\end{aligned}
\end{equation}
with $\phi_1\overleftrightarrow{\partial_\mu}\phi_2=\phi_1\partial_\mu\phi_2-(\partial_\mu\phi_1)\phi_2$.
As consequence of the trivial BRST transformations, 
one obtains trivial tree level WIs
as long as a neutral Higgs is involved:
\begin{equation}\label{eq:whh-sti}
  \begin{aligned}
&\ii p_{W,\mu}  \Gamma_{W^\pm H^\mp S}^\mu(p_W,k_\mp,k_S)\\
&\qquad \qquad+m_W\Gamma_{\phi^\pm H^\mp
  S}(p_W,k_\mp,k_S)=0\\
 &\ii p_{Z,\mu} \Gamma_{Z A^0 S}^\mu(p_Z,k_A,k_S)\\
&\qquad\qquad+m_Z\Gamma_{\phi^0 A^0
  S}(p_W,k_A,k_S)=0
 \end{aligned}
\end{equation}
Again, these WIs can be checked using the explicit form 
of the vertices~\eqref{eq:wphmh-vert}.
In the context of the flip formalism, the trivial 
WIs~\eqref{eq:whh-sti} imply the absence of the gauge flips
\begin{multline}
\parbox{28\unitlength}{
\fmfframe(2,2)(2,2){
\begin{fmfchar*}(20,15)
\fmfleft{l1,l2}
\fmfright{r1,r2}
\fmf{photon}{r1,fwf1}
\fmf{dashes}{fwf1,r2}
\fmf{dashes,label=$S,,A^0$}{fwf1,fwf2}
\fmf{plain}{fwf2,l1}
\fmf{plain}{fwf2,l2}
\fmfdot{fwf1}
\fmfdot{fwf2}
\fmfv{label=$H^\pm,,A^0,,S$,la.si=right,la.di=0.1cm}{r2}
\fmfv{label=$W^\mp,,Z$,la.si=right,la.di=0.1cm}{r1}
\end{fmfchar*}}}\quad \nLeftrightarrow \\[5mm]
\left\{
\parbox{28\unitlength}{
\fmfframe(2,5)(5,5){
\begin{fmfchar*}(15,15)
\fmfleft{l1,l2}
\fmfright{r1,r2}
\fmf{photon}{r1,fwf1}
\fmf{dashes}{fwf2,r2}
\fmf{plain}{l1,fwf1,fwf2,l2}
\fmfdot{fwf1}
\fmfdot{fwf2}
\fmfv{label=$H^\pm,,A^0,,S$,la.si=right,la.di=0.1cm}{r2}
\fmfv{label=$W^\mp,,Z$,la.si=right,la.di=0.1cm}{r1}
\end{fmfchar*}}}\quad,\quad
\parbox{35\unitlength}{
\fmfframe(0,5)(5,5){
\begin{fmfchar*}(15,15)
\fmfleft{l1,l2}
\fmfright{r1,r2}
\fmf{phantom}{r1,fwf1}
\fmf{phantom}{fwf2,r2}
\fmf{plain}{l1,fwf1,fwf2,l2}
\fmffreeze
\fmf{photon}{r1,fwf2}
\fmf{dashes,rubout}{fwf1,r2}
\fmfdot{fwf1}
\fmfdot{fwf2}
\fmfv{label=$H^\pm,,A^0,,S$,la.si=right,la.di=0.1cm}{r2}
\fmfv{label=$W^\mp,,Z$,la.si=right,la.di=0.1cm}{r1}
\end{fmfchar*}}}
\right\}
\end{multline}
Thus internal neutral Higgs bosons don't appear in the gauge flips,
and therefore can be treated exactly like the SM Higgs in a nonlinear
parameterization. This suggests the validity of the naive ET and the
gauge parameter independence of the propagators of the neutral
scalars, like in the NL-SM.  On a formal level, similarly to the
discussion in sections~\ref{sec:et} and~\ref{sec:nl}, these properties
are indeed a consequence of the trivial BRST transformations.

The gauge parameter dependence of the
self energies of the neutral Higgs bosons is governed by an identity
of the same form as~\eqref{eq:higgs-nielsen}, also for mixed two point
functions:
\begin{equation}\label{eq:2hdm-nielsen}
  \partial_\xi \Gamma_{H^0_iH^0_j}=\Lambda_{\phi^{0}H^0_i}\Gamma_{\phi^0H^0_j}+\Lambda_{ZH^0_i}^\mu\Gamma_{ZH^0_j,\mu}+\,(i\leftrightarrow j)
\end{equation}
where $H^0_i\in\{H,S,A^0\}$.
Thus the self energies can become gauge parameter dependent only if
gauge-Higgs mixing occurs. This can affect the CP-odd scalar $A^0$ already
on the one loop level. In contrast, mixing of CP even and CP
odd Higgs bosons will induce no further gauge parameter dependence since no
mixed two point functions like $\Gamma_{HA}$ appear on the right
hand side of~\eqref{eq:2hdm-nielsen}.

\subsection{Charged Higgs bosons} 
\label{sec:charged}
While the situation for neutral Higgs bosons resembles that  in the NL-SM, 
the nontrivial BRST transformation of the charged Higgs 
bosons~\eqref{eq:higgs-brst} implies that they have to be treated similar
as in a linear parameterization.
The BRST transformation~\eqref{eq:higgs-brst} implies a nontrivial tree level 
STI for the $ZH^- H^+$ vertex:
\begin{multline}\label{eq:zhpm-sti}
\ii p_{Z,\mu}  \Gamma_{ZH^+H^-}^\mu(p_Z,k_+,k_-)+m_Z\Gamma_{\phi_0H^+H^-}(p_Z,k_+,k_-)\\
=\ii\tfrac{g}{\cos\theta_w}\sin^2\theta_w\left[ D_{H^+H^-}^{-1}(k_+)-D_{H^+H^-}^{-1}(k_-)\right]
\end{multline}
The STI for the $\gamma H^+ H^-$ vertex is similar.
One contribution to the $ZH^+H^-$ vertex arises from
the operator $\mathcal{L}_{\mathcal{H}U}$:
\begin{subequations}\label{eq:zhh} 
  \begin{equation}\label{eq:zhh1}
  \mathcal{O}^{\mathcal{H}U}_{ZH^+H^-}:=\frac{\ii}{v}
(\partial^\mu\phi_0+m_Z Z^\mu)(H^+\overleftrightarrow{\partial_\mu} H^-)
  \end{equation}
  The corresponding Feynman rule arising from this operator satisfies
  a trivial tree level WI, i.e. it doesn't contribute to the right
  hand side of~\eqref{eq:zhpm-sti}.  This reflects the separate gauge
  invariance of the operator $\mathcal{L}_{\mathcal{H}U}$ that can
  appear with an arbitrary coefficient $\lambda_{\mathcal{H}U}$.
  However, there is another contribution to the $ZH^+H^-$ vertex from
  the kinetic term of the Higgs bosons $\mathcal{L}_{\mathcal{H}}$, so
  its coefficient is fixed in agreement with the
  STI~\eqref{eq:zhpm-sti}:
\begin{equation}\label{eq:zhh2}
\mathcal{O}^{\mathcal{H}}_{VH^+H^-}:=\ii(H^+\overleftrightarrow{\partial_\mu} H^-)(eA^\mu-\frac{g}{\cos\theta_w}\sin^2\theta_wZ^\mu)
\end{equation}
\end{subequations}
Because of the nontrivial STI~\eqref{eq:zhpm-sti}, diagrams with
internal charged Higgs bosons must be included in the gauge flips.
For example, the flips for sub-amplitudes including 2-fermion, charged
Higgs bosons and neutral gauge bosons are given by
\begin{multline}
{\tilde G}_{4,1H^\pm 2F}\\
=\left\{
\parbox{20\unitlength}{
\fmfframe(0,2)(2,2){
\begin{fmfchar*}(15,15)
\fmfleft{a,b}
\fmfright{f1,f2}
\fmf{fermion}{a,fwf1}
\fmf{fermion}{fwf1,fwf2}
\fmf{fermion}{fwf2,b}
\fmf{photon}{fwf1,f1}
\fmf{dashes}{fwf2,f2}
\fmfv{label= $Z,,\gamma$,la.di=0.1cm}{f1}
\fmfv{label= $H^\pm$ ,la.di=0.1cm}{f2}
\fmfdot{fwf1,fwf2}
\end{fmfchar*}}}\, ,\,
\parbox{20\unitlength}{
\fmfframe(2,2)(2,2){
\begin{fmfchar*}(15,15)
\fmfleft{a,b}
\fmfright{f1,f2}
\fmf{fermion}{a,fwf1}
\fmf{fermion}{fwf1,fwf2}
\fmf{fermion}{fwf2,b}
\fmf{phantom}{fwf2,f2}
\fmf{phantom}{fwf1,f1}
\fmffreeze
\fmf{photon}{fwf2,f1}
\fmf{dashes}{fwf1,f2}
\fmfdot{fwf1,fwf2}
\fmfv{label= $Z,,\gamma$,la.di=0.1cm}{f1}
\fmfv{label= $H^\pm$ ,la.di=0.1cm}{f2}
\end{fmfchar*}}}\, ,\, 
\parbox{25\unitlength}{
\fmfframe(2,2)(5,2){
\begin{fmfchar*}(15,15)
\fmfleft{a,b}
\fmfright{f1,f2}
\fmf{fermion}{a,fwf}
\fmf{fermion}{fwf,b}
\fmf{dashes,label= $H^\pm$}{fwf,www}
\fmf{photon}{www,f1}
\fmf{dashes}{www,f2}
\fmfdot{www,fwf}
\fmfv{label= $Z,,\gamma$,la.di=0.1cm}{f1}
\fmfv{label= $H^\pm$ ,la.di=0.1cm}{f2}
\end{fmfchar*}}}
\right \} 
\end{multline}
both in the linear and the nonlinear parameterization.

As a consequence,  there is 
a flip connecting self-energies and vertex corrections, as can be seen in 
the  example of  the coupling of a charged
Higgs boson to $b$ and $t$ quarks:
 \begin{equation}
\parbox{30mm}{
\begin{fmfgraph*}(30,20)
\fmfleft{l1,l2}
\fmfright{r1,r2}
\fmf{fermion,la=$t$,la.si=left}{l1,v}
\fmf{fermion,la=$\bar b$}{v,l2}
\fmf{plain}{r1,r,r2}
\fmf{dashes,tension=0.5,label=$H^+$}{v,i1}
\fmf{photon,left,tension=0.25,label=$Z$,la.si=left}{i1,i2}
\fmf{dashes,left,tension=0.25,label=$H^+$,la.si=left}{i2,i1}
\fmf{dashes,tension=0.5,label=$H^+$}{i2,r}
\fmfdot{v,r,i1,i2}
\end{fmfgraph*}}\quad\Leftrightarrow\quad
\parbox{30mm}{
\begin{fmfchar*}(30,20)
\fmfleft{l1,l2}
\fmfright{r1,r2}
\fmf{fermion,label=$t$,tension=2,la.si=left}{l1,i1}
\fmf{fermion,label=$\bar b$,tension=2,la.si=left}{i2,l2}
\fmf{plain}{r1,r,r2}
\fmf{dashes,label=$H^+$,tension=1,la.si=right}{i1,v}
\fmf{photon,label=$Z$,label.si=right}{v,i2}
\fmf{dashes,label=$H^+$,la.si=left}{v,r}
\fmffreeze
\fmf{fermion,tension=0,label=$\bar b$,la.si=left}{i1,i2}
\fmfdot{v,r,i1,i2}
\end{fmfchar*}}
\end{equation}
Also, there are flips to irreducible 
higher loop contributions to the self energy:
\begin{equation}
\parbox{40mm}{
\begin{fmfgraph*}(40,20)
\fmfleft{l1,l2}
\fmfright{r1,r2}
\fmf{plain}{l1,v,l2}
\fmf{plain}{r1,r,r2}
\fmf{dashes,tension=0.5,label=$H^+$}{v,i1}
\fmf{photon,left,tension=0.2}{i1,i2}
\fmf{dashes,left,tension=0.2}{i2,i1}
\fmf{dashes,tension=0.5,label=$H^+$}{i2,i3}
\fmf{photon,left,tension=0.2}{i3,i4}
\fmf{dashes,left,tension=0.2}{i4,i3}
\fmf{dashes,tension=0.5,label=$H^+$}{i4,r}
\fmfdot{v,r,i1,i2,i3,i4}
\end{fmfgraph*}}
\,\Leftrightarrow\,
\parbox{35mm}{
\begin{fmfgraph*}(35,20)
\fmfleft{l1,l2}
\fmfright{r1,r2}
\fmftop{t}
\fmfbottom{b}
\fmf{plain}{l1,v,l2}
\fmf{plain}{r1,r,r2}
\fmf{phantom,tension=3}{t,il1}
\fmf{dashes,tension=1}{il1,il2}
\fmf{phantom,tension=3}{il2,b}
\fmf{dashes,tension=0.5,label=$H^+$}{v,i1}
\fmf{photon,tension=0.2}{i1,il1}
\fmf{dashes,tension=0.2}{il1,i2}
\fmf{dashes,tension=0.2}{il2,i1}
\fmf{photon,tension=0.2}{i2,il2}
\fmf{dashes,tension=0.5,label=$H^+$}{i2,r}
\fmfdot{v,r,i1,il1,i2,il2}
\end{fmfgraph*}}
\end{equation}
This indicates that charged Higgs bosons in the 2HDM cannot be resummed 
consistently, also in a nonlinear parameterization.
In agreement with this result, it is shown in appendix~\ref{app:nielsen}
that the self energy of the charged Higgses is
in general gauge parameter dependent, also without taking 
mixing with the gauge sector into account.

One also expects that the naive version of the ET is violated 
for intermediate charged Higgs bosons, i.e. for a finite width of the 
charged Higgs
the gauge boson amplitudes  and the GB amplitudes don't agree manifestly.
As an explicit example, 
consider the  process $Z\to t \bar bH^-$ that appears 
as subprocess for associated production of charged Higgs bosons at
linear colliders.
In the nonlinear 2HDM, Yukawa couplings can be obtained from the operator
\begin{equation}\label{eq:2hdm-yukawa}
  \mathcal{L}_Y=-\bar Q_L U\mathcal{H}
  \left[\tfrac{(\lambda_t+\lambda_b)}{2}+\tfrac{(\lambda_t-\lambda_b)}{2}\sigma^3\right]Q_R
  +\text{h.c.}
\end{equation}
with $m_b=\lambda_b v_1=\lambda_b v\cos\beta$ and  $m_t=\lambda_t v_2=\lambda_t v\sin\beta$. 
The choice~\eqref{eq:2hdm-yukawa} corresponds to the so called type 
II 2HDM~\cite{Carena:2002es}.
From an effective field theory perspective, an additional term involving only the GB matrix
can be added to~\eqref{eq:2hdm-yukawa}, as in the first term 
of~\eqref{eq:nl-yukawa}. The effects of such nonstandard
Yukawa couplings on unitarity have been discussed in
section~\ref{sec:et} and will not be considered in the following.
The resulting Yukawa couplings of the charged 
Higgs bosons are
\begin{multline}\label{eq:charged-yukawa}
  \mathcal{L}_{Y H^\pm}=
-\frac{\sqrt 2}{v}\left(1+\ii\frac{\phi^0}{v}\right)
  \Bigl[\bar t H^+\bigl(m_b\tan\beta\left(\tfrac{1+\gamma^5}{2}\right)\\
      + m_t\cot\beta\left(\tfrac{1-\gamma^5}{2}\right)\bigr)b\Bigr]
 + \text{h.c.}+\dots
\end{multline}

In the computation of the diagrams, we consider again 
the limit\footnote{For simplicity, we assume in addition 
$m_b\tan\beta \ll m_t$ but this is not essential to our argument.} 
$m_b,m_Z\to 0$.
We will also include an arbitrary coefficient $\lambda_{\mathcal{H}U}$
as factor in front of~\eqref{eq:zhh1}.
Similar to the example of top pair production in section~\ref{sec:et}, 
the only GB diagrams with dangerous high energy behavior are the 
Higgs exchange and the  contact diagram. 
Using the Feynman rules from~\eqref{eq:zhh1}
 and~\eqref{eq:charged-yukawa} we obtain
\begin{multline}\label{eq:charged-et-gb}
\parbox{25mm}{
 \fmfframe(2,5)(5,5){
\begin{fmfchar*}(15,15)
\fmfleft{f1}
\fmfright{a,b,f2}
\fmf{fermion}{a,fwf}
\fmf{fermion}{fwf,b}
\fmf{dashes,tension=2,label=$\phi^0$}{fwf,f1}
\fmf{dashes}{fwf,f2}
\fmfdot{fwf}
\fmfv{label= $H^-$ ,la.di=0.2cm}{f2}
\fmfv{label= $\bar b$}{a}
\fmfv{label= $t$ ,la.di=0.2cm}{b}
\end{fmfchar*}}}+
\parbox{25mm}{
 \fmfframe(0,5)(5,5){
\begin{fmfchar*}(20,15)
\fmfleft{f1}
\fmfright{a,b,f2}
\fmf{dashes,tension=2,label=$\phi^0$}{www,f1}
\fmf{phantom}{www,a}
\fmf{phantom}{www,f2}
\fmffreeze
\fmf{fermion}{a,fwf}
\fmf{fermion}{fwf,b}
\fmf{dashes,label= $H^+$,la.si=left}{fwf,www}
\fmf{dashes}{www,f2}
\fmfdot{www,fwf}
\fmfv{label= $H^-$ ,la.di=0.2cm}{f2}
\fmfv{label= $\bar b$ ,la.di=0.2cm}{a}
\fmfv{label= $t$ ,la.di=0.2cm}{b}
\end{fmfchar*}}}\\
= \frac{\sqrt 2 m_t\cot\beta}{v^2}
\left[\bar t\left(\tfrac{1-\gamma^5}{2}\right)b\right]
\Biggl(1-\\\lambda_{\mathcal{H}U}\frac{(p_{H^+}^2-p_{H^-}^2)}{p_{H^+}^2-m^2_{H^+}+\ii\Im\Pi_{H^+}(p_{H^+}^2)}\Biggr)
\end{multline}
In the case of a standard coefficient
of~\eqref{eq:zhh2}, i.e. $\lambda_{\mathcal{H}U}=1$, the
expression~\eqref{eq:charged-et-gb} vanishes if the external $H^-$ is
on its mass shell ($p_{H^-}^2=m_{H^+}^2$) and the width is set to
zero\footnote{ Actually, this could have been anticipated since in a
  linear parameterization the coupling of a GB to two particles
  with the same mass vanishes (see e.g. the second
  reference in~\cite{OS:FGH}) so no diagram of the
  form~\eqref{eq:charged-et-gb} appears at all.}. Since we cannot
expect the ET to hold for an off-shell $H^-$ (see below) and in order
to make the cancellations among different diagrams more transparent, in
the following we will keep $\lambda_{\mathcal{H}U}\neq 1$.
 
In the example of section~\ref{sec:et-example}, we were able to
decompose the amplitude into two groves that satisfied the ET by
themselves.  This is not the case in the present example and all
diagrams for longitudinal gauge bosons are needed to reproduce the GB
amplitude~\eqref{eq:charged-et-gb}, as we will show now.  The diagrams
with internal fermions give an additional term compared to the contact
term in~\eqref{eq:charged-et-gb}:
\begin{multline}\label{eq:charged-et-compton}
\parbox{30\unitlength}{
\fmfframe(5,5)(5,5){
\begin{fmfchar*}(20,15)
\fmfleft{f1}
\fmfright{a,b,f2}
\fmf{photon,tension=2,label= $Z_L$}{fwf1,f1}
\fmf{fermion,tension=2}{fwf1,fwf2}
\fmf{fermion,tension=2}{fwf2,f2}
\fmf{fermion,tension=1}{a,fwf1}
\fmffreeze
\fmf{dashes}{fwf2,b}
\fmfv{label= $H^-$ ,la.di=0.05cm}{b}
\fmfv{label= $\bar b$ ,la.di=0.1cm}{a}
\fmfv{label= $t$ ,la.di=0.1cm}{f2}
\fmfdot{fwf1,fwf2}
\end{fmfchar*}}}\, +
\parbox{30\unitlength}{
\fmfframe(2,5)(5,5){
\begin{fmfchar*}(20,15)
\fmfleft{f1}
\fmfright{a,b,f2}
\fmf{photon,tension=2,label= $Z_L$}{fwf1,f1}
\fmf{fermion,tension=2}{fwf2,fwf1}
\fmf{fermion,tension=2}{a,fwf2}
\fmf{fermion,tension=1}{fwf1,f2}
\fmffreeze
\fmf{dashes}{fwf2,b}
\fmfv{label= $H^-$ ,la.di=0.1}{b}
\fmfv{label= $\bar b$ ,la.di=1}{a}
\fmfv{label= $t$ ,la.di=1}{f2}
\fmfdot{fwf1,fwf2}
\end{fmfchar*}}}
\\=-\frac{\ii g\sqrt 2 m_t\cot\beta}{v\cos\theta_w m_Z}(\tfrac{1}{2}-\sin^2\theta_w)
\left[\bar t\left(\tfrac{1-\gamma^5}{2}\right)b\right]+\dots
\end{multline}
where we have suppressed the terms corresponding to the 
GB diagrams with internal fermion lines we have omitted above.
The Higgs exchange diagram gives, using the Feynman rules obtained from
both operators~\eqref{eq:zhh} 
\begin{multline}\label{eq:charged-et-higgs}
\parbox{25mm}{
 \fmfframe(0,5)(5,5){
\begin{fmfchar*}(20,15)
\fmfleft{f1}
\fmfright{a,b,f2}
\fmf{photon,tension=2,label=$Z_L$}{www,f1}
\fmf{phantom}{www,a}
\fmf{phantom}{www,f2}
\fmffreeze
\fmf{fermion}{a,fwf}
\fmf{fermion}{fwf,b}
\fmf{dashes,label= $H^+$,la.si=left}{fwf,www}
\fmf{dashes}{www,f2}
\fmfdot{www,fwf}
\fmfv{label= $H^-$ ,la.di=0.2cm}{f2}
\fmfv{label= $\bar b$ ,la.di=0.2cm}{a}
\fmfv{label= $t$ ,la.di=0.2cm}{b}
\end{fmfchar*}}}=
-\frac{\ii g\sqrt 2 m_t\cot\beta}{v\cos\theta_w m_Z}
\left(\sin^2\theta_w-\tfrac{\lambda_{\mathcal{H}U}}{2}\right)\\
\left[\bar t\left(\tfrac{1-\gamma^5}{2}\right)b\right]\frac{(p_{H^+}^2-p^2_{H^-}) }{p_{H^+}^2 -m^2_{H^+}+\ii\Im\Pi_{H^+}(p_{H^+}^2)}
\end{multline}
Only for a vanishing width and $H^-$ on the mass shell, the terms
proportional to $\sin^2\theta_w$ cancel
between~\eqref{eq:charged-et-compton} and~\eqref{eq:charged-et-higgs}
and the GB amplitude~\eqref{eq:charged-et-gb} is reproduced (up to
a phase).  Therefore the situation for the charged Higgs in the
nonlinear parameterization is similar to a linear parameterization,
as described in subsection~\ref{sec:et-example}, and the naive version of
the ET is not satisfied when a Dyson summation of the charged Higgs
propagator is performed.  Also since the external charged Higgs must be on
the mass shell, an ET for off-shell Higgs bosons~\eqref{eq:higgs-et}
is not valid for the charged Higgs bosons.  These results therefore
confirm the expectation of the flip formalism.
\section{Summary and outlook}
Motivated by the structure of gauge invariant classes of tree diagrams
in nonlinear parameterizations of the scalar sector~\cite{OS:FGH} 
and the observations of~\cite{Valencia:1990jp} concerning 
effects of the Higgs width on the Goldstone
boson equivalence theorem, we have
revisited the properties of the Higgs resonance in nonlinear 
parameterizations.
As we have demonstrated for the nonlinear parameterizations of both the
minimal standard model and a two-Higgs doublet model, the Dyson 
summation of propagators of neutral Higgs bosons can be performed 
 without violating gauge parameter independence  and Ward identities.
Although in nonlinear parameterizations care must be taken not to violate 
bounds from tree unitarity, a simple unitarity restoring
expression for the Higgs propagator~\cite{Seymour:1995qg}
can be used without violating the naive equivalence theorem, in contrast to
linear parameterizations.

Furthermore, the full Higgs propagator has been shown to be
gauge parameter independent. For the Higgs self 
energy this holds only in the absence of CP violating mixing 
with the gauge sector.
These results are consistent with the conjectured extension of the
`gauge flip' formalism to loop diagrams~\cite{Boos:1999}.

For charged Higgs bosons in a two-Higgs doublet model, gauge flips
exist that connect resummed self energy diagrams to irreducible higher
order contributions to the self energy or to vertex corrections and a
Dyson summation is not compatible with gauge invariance.  The
violation of the naive equivalence theorem has been demonstrated for
an explicit example.

The Higgs resonance has served as a first example of the application
of the flip formalism to loop diagrams in a case where independent
methods are available to verify the results.  As mentioned in
section~\ref{sec:groves}, a second example where the flips reproduce
results established by different methods is the fermion loop
scheme~\cite{Argyres:1995}.  A formal proof of the
gauge flip formalism for one loop diagrams in linear parameterizations
and applications to one loop SM processes with up to 4 fermions in the
final state will be given elsewhere~\cite{Ondreka:2003,OOS:prep}.  We
hope the formalism will prove useful in situations where direct proofs
are difficult to achieve.  
\section*{Acknowledgements}
I thank Thorsten Ohl and David Ondreka for many useful 
discussions and Hubert Spiesberger for helpful comments on the manuscript. 
This work has been supported
by the Deutsche
Forschungsgemeinschaft through the Gra\-du\-ier\-ten\-kolleg `Eichtheorien' at
Mainz University.
\appendix 
\section{Slavnov-Taylor and Nielsen Identities}
In this appendix we give some technical details on formulae used in the main
text and set up our notation for the functional identities resulting
from gauge invariance.
\subsection{Zinn-Justin Identity and STIs}\label{app:sti}
The derivation of STIs for irreducible vertices uses the Zinn-Justin
Identity
\begin{equation}\label{eq:sti}
\sum_\Psi\int\dd^4 x \frac{\delta\Gamma}{\delta \Psi^\star}\frac{\delta
  \Gamma}{\delta \Psi}+B_a\frac{\delta\Gamma}{\delta\bar c_a}=0\,.
\end{equation}
where the $\Psi$ summarize all fields in the theory and the $\Psi^\star$ 
are the sources of the BRST transformations included in  the effective action
$ \Gamma=\Gamma_0+\sum_\Psi\int\dd^4x\,\tr\lbrack {\Psi^\star}
 (\delta_{\text{BRS}} \Psi)\rbrack $. We will use the notation
$\Gamma_{\Psi_1\dots\Psi_n}=\frac{\delta^n\Gamma}{\delta\Psi_1\dots\delta\Psi_n}|_{\Psi=0}$ for the irreducible vertex functions.
From~\eqref{eq:sti} we obtain the general
 STI for the $HVV$ and the $H\phi V$ vertices
by taking a derivative with respect to a ghost, a Higgs field and a gauge
boson or a GB:
\begin{subequations}\label{eq:h-stis}
\begin{align}
\label{eq:hww-sti}
&-\sum_{\Psi=V_a,\phi_a}\Gamma_{c_{a} \Psi^\star}
\Gamma_{\Psi V_b H}=\\\nonumber
&\sum_{\Psi=V_a,\phi_a,H}\left(
  \Gamma_{c_{a} \Psi^\star V_b}\Gamma_{\Psi H} 
 +\Gamma_{c_{a} \Psi^\star H}\Gamma_{\Psi V_b}\right)
  + \tfrac{\ii}{\xi}p_b^\nu \Gamma_{c_a\bar c_b H}\\
\label{eq:hphiw-sti}
&-\sum_{\Psi=V_a,\phi_a}\Gamma_{c_{a} \Psi^\star}
\Gamma_{\Psi \phi_b H}=\\ \nonumber
&\sum_{\Psi=V_a,\phi_a,H}\left(
  \Gamma_{c_{a} \Psi^\star \phi_b}\Gamma_{\Psi H}
+ \Gamma_{c_{a} \Psi^\star H}\Gamma_{\Psi \phi_b}\right)
  + m_{W_a} \Gamma_{c_a\bar c_b H}
\end{align}
\end{subequations}
where we have used the equation of motion for the auxiliary field $B$.
For a nonlinear transformation law of the GBs, 
there are additional contributions of the
form $\Gamma_{c\phi^\star\phi\phi }\Gamma_{H}$ that vanish 
in the absence of tadpoles. In this case both relations~\eqref{eq:h-stis} are
valid both for the linear
and nonlinear parameterizations.
In the nonlinear parameterization the Higgs drops out of the sums on
the right hand side due to its trivial BRST transformation so 
 no Higgs two point functions appear in these STIs.
In higher orders, the Higgs ghost couplings $\Gamma_{c_a\bar c_b H}$
and the vertex functions $\Gamma_{c_{a} \phi_b^\star H}$  and $\Gamma_{c_{a} V
_b^\star H}$ 
can be generated radiatively
but they are absent on tree level in the nonlinear parameterization. 
Using the linear terms in 
the BRST transformations of the gauge bosons  $\delta_{\text{BRST}} V_a=\partial_\mu c_a+\dots$ and 
GBs $\delta_{\text{BRST}}\phi_a=-m_{V_a}c_a+\dots$ 
we thus arrive at the 
simple tree level WIs~\eqref{eq:h-wis}.
\subsection{Gauge parameter dependence of Green's functions and irreducible vertices}
\label{app:nielsen}
An identity for the gauge parameter dependence of irreducible
vertices can be derived using an extended  BRST 
symmetry~\cite{Piguet:1985}.  For simplicity, 
we suppress the indices distinguishing the gauge bosons in the following.
Introducing an auxiliary Grassmann variable $\chi$ allows to give  
the gauge parameters themselves a transformation law
\begin{equation}
\delta_{\text{BRST}}\, \xi =\chi \quad,\quad \delta_{\text{BRST}}\,\chi=0
\end{equation}
In the usual BRST formalism, 
the gauge fixing and ghost Lagrangian is a BRST exact operator, i.e.
it can be written as a BRST transformation of a functional $\Theta$ 
of ghost number $(-1)$.
In the extended BRST formalism 
we get an additional contribution from the transformation of $\xi$ so that  
\begin{equation}
\delta_{\text{BRST}}\Theta=\mathcal{L}_{GF}+\mathcal{L}_{FP}+\mathcal{L}_{\chi}
\text{ with  }\mathcal{L}_{\chi}=\chi\partial_{\xi}\Theta  
\end{equation}
For the usual $R_\xi$ gauge fixing 
$ \Theta=\bar c(\partial_\mu V^\mu-\xi m_{V}\phi +\frac{\xi}{2}B)$
we find
\begin{equation}\label{eq:chi-lag}
  \begin{aligned}
 \mathcal{L}_{\chi}=\frac{1}{2}\chi\bar c (B-2 m_{V}\phi)
   = -\frac{1}{2\xi}\chi\bar c (\partial_\mu V^\mu+\xi m_{V}\phi)
  \end{aligned}
\end{equation}
where in the last step we have used the equation of motion 
for the auxiliary field $B$.

The Zinn-Justin identity resulting from the extended 
BRST transformation is given by
\begin{equation}\label{eq:ext-zje}
\sum_\Psi\int \dd^4 x \frac{\delta \Gamma}{\delta \Psi^\star}\frac{\delta
  \Gamma}{\delta \Psi}+B\frac{\delta \Gamma}{\delta\bar c}+\chi \partial_\xi\Gamma=0
\end{equation}
Taking the derivative with respect to $\chi$, taking the fermionic character
into account,  
one obtains the so called `Nielsen Identity'~\cite{Piguet:1985}
\begin{equation}\label{eq:nielsen}
 \partial_\xi\Gamma
=\sum_\Psi\int \dd^4 x \frac{\delta \Gamma}{\delta \Psi^\star}\frac{\delta
 \Gamma_\chi}{\delta \Psi}+\frac{\delta\Gamma_\chi }{\delta \Psi^\star}\frac{\delta \Gamma}{\delta \Psi}  +B\frac{\delta\Gamma_\chi }{\delta\bar c}
\end{equation}
where we have introduced the notation $\Gamma_\chi=\partial_\chi\Gamma |_{\chi=0}$.
The vertices involving insertions of $\chi$ can be evaluated using the
 lagrangian~\eqref{eq:chi-lag}.
The renormalization conditions of physical
parameters have to be chosen appropriately
so the Nielsen Identity~\eqref{eq:nielsen} is not deformed
in higher orders~\cite{Gambino:1999}.

As discussed in~\cite{Gambino:1999}, imposing the vanishing of the Higgs
tadpole as renormalization condition implies $\Gamma_{\chi H^*}= 
\Gamma_{\chi \phi^{0*}}=0$.
Taking the derivative
of the Nielsen Identity~\eqref{eq:nielsen} with respect to  $Z$
we get a relation connecting these functions to  $\Gamma_{\chi Z^*}$ 
 so it is also constrained to vanish:
\begin{equation}
  \Gamma_{\chi Z^*}\Gamma_{ZZ}+\Gamma_{\chi \phi^{0*}}\Gamma_{\phi^0 Z}+
\Gamma_{\chi H^{*}}\Gamma_{H Z}=0
\end{equation}
In the nonlinear parameterization, $\delta_{\text{BRST}} H=0$ implies there are no
 vertex functions involving $H^*$ and one
obtains for the self energies of the neutral sector
\begin{multline}\label{eq:neutral-nielsen}
   \partial_\xi \Gamma_{H\Phi}=\Gamma_{\chi
  \phi^{0*}H}\Gamma_{\phi^0\Phi}+\Gamma_{\chi Z^*H}^\mu\Gamma_{Z\Phi,\mu}\\
+\Gamma_{\chi
  \phi^{0*}\Phi}\Gamma_{\phi^0H}+\Gamma_{\chi Z^*\Phi}^\mu\Gamma_{Z H,\mu}
\end{multline}
with $\Phi\in\{H,\phi^0,Z\}$. 
For the Higgs self energy we obtain~\eqref{eq:higgs-nielsen} with
$\Lambda_{\Phi H}=\Gamma_{\chi \Phi^* H}$.
The important thing is the absence
of terms involving the Higgs two point function  on the right hand
side, while the mixing
of the Higgs is due only to CP violation. Therefore in the 
absence of CP violation one has $\partial_\xi \Gamma_{HH}=0$.
In the 2HDM, a similar identity holds for all neutral Higgs bosons (including
the CP odd) since they transform trivially.
For the self energy of the charged Higgs boson one has instead
\begin{multline}
   \partial_\xi \Gamma_{H^+H^-}=
  \Gamma_{\chi
  H^{+*}H^-}\Gamma_{H^-H^+}
  +\Gamma_{\chi\phi^{+*}H^-}\Gamma_{\phi^-H^+}\\
  +\Gamma_{\chi W^{+*}H^-}\Gamma_{W^-_LH^+}
   +\,(+\leftrightarrow -)   
\end{multline}
Here the self energy itself appears on the right hand side and it will in
general be gauge parameter dependent.

The identity governing the gauge parameter dependence of  Green's 
functions can be derived from the extended Zinn-Justin 
Identity~\eqref{eq:ext-zje}
by Legendre transformation~\cite{Piguet:1985}
or directly from the path integral representation of Green's functions.
In operator language, it is given by
\begin{multline}\label{eq:nielsen-gf}
\partial_\xi\Greensfunc{\Psi_1\dots\Psi_n}=\\
-\sum_{\Psi_i}\pm \Greensfunc{\left
(\ii\int \dd^4 x \,\partial_\xi\Theta\right) 
\Psi_1\dots \delta_{\text{BRST}} \Psi_i\dots \Psi_n}
\end{multline}
where the signs arise for fermionic fields anticommuting with the 
BRST transformation. 
Applying the LSZ formula to a given field in the Green's function, 
the contribution from the BRST 
transformed fields factorizes and can be absorbed in the wave function 
renormalization so~\eqref{eq:nielsen-gf} is also valid if the 
external vacuum states
are replaced by physical $|\text{in}\rangle$ and $|\text{out}\rangle$ states. 

\providecommand{\href}[2]{#2}

\end{fmffile}
\end{document}